\documentclass[12pt]{article}

\usepackage{verbatim,color,amssymb,enumerate}
\usepackage{amsmath}					
\usepackage{afterpage}					
\usepackage{amsthm}					
\usepackage{natbib}
\usepackage{multirow}
\usepackage{setspace}
\usepackage[mathscr]{euscript}
\usepackage{fancyhdr}
\usepackage{enumitem}
\usepackage{graphicx}
\usepackage{geometry}
\usepackage[bookmarks=false]{hyperref}
\usepackage{algorithm, algorithmicx}
\usepackage[normalem]{ulem}

\usepackage{multibib}
\newcites{latex}{References}

\usepackage{tabularx, booktabs}
\newcolumntype{Y}{>{\centering\arraybackslash}X}
\usepackage{lscape}

\usepackage{subfig}
\usepackage{float}
\usepackage{lineno}

\newtheorem{Thm}{\underline{\bf Theorem}}
\newtheorem{Lem}{\underline{\bf Lemma}}

\def\eE{\mathbb{E}}

\def\S{{\cal S}}

\def\diag{\hbox{diag}}

\def\diag{\hbox{diag}}

\def\Dir{\hbox{Dir}}

\def\HC{\hbox{C}^{+}}

\def\MVN{\hbox{MVN}}

\def\Normal{\hbox{Normal}}

\def\Mult{\hbox{Mult}}

\def\P_25_ICML{{\it Proceedings of the 25th international conference on Machine learning}}

\def\bse{\begin{eqnarray*}}
\def\ese{\end{eqnarray*}}
\def\be{\begin{eqnarray}}
\def\ee{\end{eqnarray}}
\def\bq{\begin{equation}}
\def\eq{\end{equation}}

\def\trans{^{\rm T}}

\def\th{^{th}}

\def\bb{{\mathbf b}}
\def\bB{{\mathbf B}}

\def\bd{{\mathbf d}}
\def\bD{{\mathbf D}}
\def\b1e{{\mathbf e}}
\def\b1f{{\mathbf f}}

\def\bI{{\mathbf I}}

\def\bM{{\mathbf M}}

\def\bP{{\mathbf P}}

\def\bs{{\mathbf s}}

\def\bw{{\mathbf w}}

\def\bz{{\mathbf z}}

\def\bzero{{\mathbf 0}}

\newcommand{\bmu}{\mbox{\boldmath $\mu$}}
\newcommand{\bnu}{\mbox{\boldmath $\nu$}}

\newcommand{\bpi}{\mbox{\boldmath $\pi$}}

\newcommand{\btheta}{\mbox{\boldmath $\theta$}}

\newcommand{\bbeta}{\mbox{\boldmath $\beta$}}
\newcommand{\bgamma}{\mbox{\boldmath $\gamma$}}
\newcommand{\bzeta}{\mbox{\boldmath $\zeta$}}

\newcommand{\bSigma}{\mbox{\boldmath $\Sigma$}}

\newcommand{\bGamma}{\mbox{\boldmath $\Gamma$}}
\newcommand{\btau}{\mbox{\boldmath $\tau$}}

\renewcommand\footnoterule{\kern-3pt \hrule \textwidth 2in \kern 2.6pt}

\newcommand\Algphase[1]{%
\vspace*{-.5\baselineskip}\Statex\hspace*{\dimexpr-\algorithmicindent-2pt\relax}\rule{\textwidth}{0.4pt}%
\Statex\hspace*{-\algorithmicindent}\textbf{#1}%
\vspace*{-.5\baselineskip}\Statex\hspace*{\dimexpr-\algorithmicindent-2pt\relax}\rule{\textwidth}{0.4pt}%
}

\def\boxit#1{\vbox{\hrule\hbox{\vrule\kern6pt \vbox{\kern6pt \textcolor{blue}{#1}\kern6pt}\kern6pt\vrule}\hrule}}

\def\authorfootnote#1{{\let\thefootnote\relax\footnotetext{#1}}}

\allowdisplaybreaks

\begin{document}
\thispagestyle{empty}
\baselineskip=28pt

\begin{center}
{\LARGE{\bf Bayesian Semiparametric Longitudinal\\ 
\vskip -8pt Inverse-Probit Mixed Models\\ 
for Category Learning
}}
\end{center}
\baselineskip=12pt
\vskip 5pt

%\iffalse
\begin{center}
Minerva Mukhopadhyay$^{1}$(minervam@iitk.ac.in)\\
Jacie R. McHaney$^{2}$ (j.mchaney@pitt.edu)\\
Bharath Chandrasekaran$^{2}$(b.chandra@pitt.edu)\\
Abhra Sarkar$^{3}$ (abhra.sarkar@utexas.edu)\\

\vskip 5mm 
$^{1}$Department of of Mathematics and Statistics,\\
Indian Institute of Technology,\\ 
Kanpur, UP 208016, India\\
\vskip 8pt 
$^{2}$Department of Communication Science and Disorders,\\ 
University of Pittsburgh,\\
4028 Forbes Tower, Pittsburgh, PA 15260, USA\\
\vskip 8pt 
$^{3}$Department of Statistics and Data Sciences,\\
University of Texas at Austin,\\ 
105 East 24th Street D9800, Austin, TX 78712, USA\\
\end{center}
%\fi

\vskip 5pt 
\begin{center}
{\Large{\bf Abstract}} 
\end{center}
\baselineskip=12pt

Understanding how the adult human brain learns novel categories is an important problem in neuroscience. 
Drift-diffusion models are popular in such contexts for their ability to mimic the underlying neural mechanisms. 
One such model for gradual longitudinal learning was recently developed in \cite{paulon2021bayes}. 
Fitting conventional drift-diffusion models, however, requires data on both category responses and associated response times. 
In practice, category response accuracies are often the only reliable measure recorded by behavioral scientists to describe human learning. 
To our knowledge, however, drift-diffusion models for such scenarios have never been considered in the literature before.
To address this gap, in this article,
we build carefully on 
\cite{paulon2021bayes}, 
but now with latent response times integrated out, 
to derive a novel biologically interpretable class of `inverse-probit' categorical probability models for observed categories alone. 
However, this new marginal model presents significant identifiability and inferential challenges not encountered originally for the joint model in \cite{paulon2021bayes}. %-  we are now attempting to recover the model parameters only from data on response categories. 
We address these new challenges using a novel projection-based approach with a symmetry-preserving identifiability constraint that allows us to work with conjugate priors in an unconstrained space. 
We adapt the model for group and individual-level inference in longitudinal settings. 
Building again on the model's latent variable representation, 
we design an efficient Markov chain Monte Carlo algorithm for posterior computation. 
We evaluate the empirical performance of the method through simulation experiments. 
The practical efficacy of the method is illustrated in applications to longitudinal tone learning studies.

\vskip 20pt 
\baselineskip=12pt
\noindent\underline{\bf Key Words}: 
Category learning, 
B-splines, 
Drift-diffusion models, 
Functional models, 
Inverse Gaussian distributions, 
Longitudinal mixed models, 
Speech learning

\par\medskip\noindent
\underline{\bf Short/Running Title}: Longitudinal Inverse-Probit Mixed Models

\par\medskip\noindent
\underline{\bf Corresponding Author}: Abhra Sarkar (abhra.sarkar@utexas.edu)

\pagenumbering{arabic}
\setcounter{page}{0}
\newlength{\gnat}
\setlength{\gnat}{16pt}
\baselineskip=\gnat

\newpage

\section{Introduction}

\paragraph{Scientific background.} 
\paragraph{Scientific background.} 
Categorization decisions are important in almost all aspects of our lives 
- whether it is a friend or a foe, edible or non-edible, the word /bat/ or /hat/, etc. 
The underlying cognitive dynamics are being actively studied through extensive ongoing research \citep{smith2004psychology, heekeren2004general, gold2007neural, schall2001neural, purcell2013neural, glimcher2013neuroeconomics}.

In typical multi-category decision tasks, the brain accumulates sensory evidence in order to make a categorical decision. 
This accumulation process is reflected in the increasing firing rates at local neural populations associated with different decisions.
A decision is taken when neural activity in one of these populations reaches a particular threshold level. 
The decision category that is finally chosen is the one whose decision threshold is crossed first \citep{gold2007neural,brody2016neural}. 
Changes in evidence accumulation rates and decision thresholds can be induced by differences in task difficulty and/or cognitive function 
\citep{cavanagh2011subthalamic,ding2013basal}. 
Decision-making is also regulated by demands on both the speed and accuracy of the task \citep{bogacz2010neural,milosavljevic2010drift}.

Understanding the brain activity patterns for 
different decision alternatives %, including their similarities and differences, 
is a key scientific interest in modeling 
brain mechanisms underlying decision-making. 
Statistical approaches with biologically interpretable parameters 
that further allow probabilistic clustering of the parameters \citep{lau2007bayesian,wade2023cluster} associated with different competing choices 
can facilitate such inference, 
the parameters clustering together indicating 
similar behavior and difficulty levels.

\begin{figure}[!ht]
	\centering
	\subfloat[Drift-diffusion processes for tone learning when data on both response categories and response times are available.]{
		\includegraphics[width=0.645\linewidth]{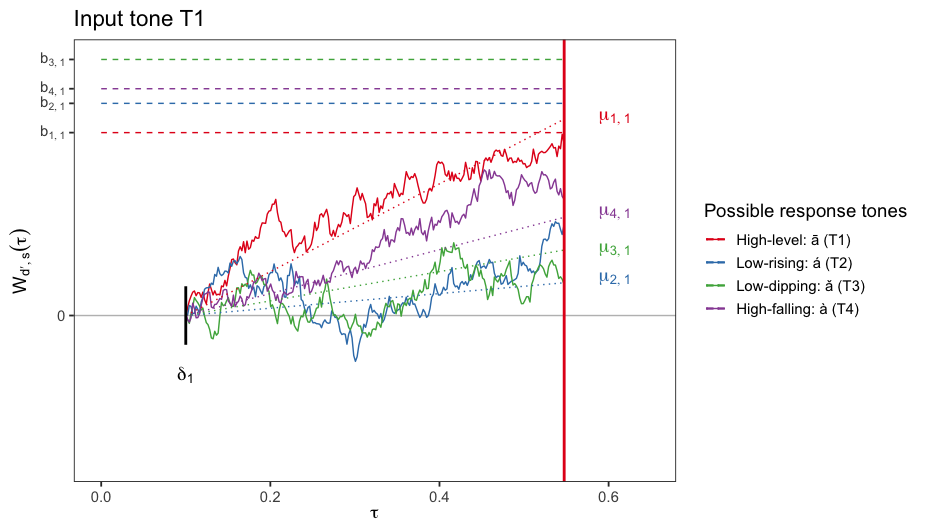}
	}\qquad
	\subfloat[Drift-diffusion processes for tone learning proposed in this article 
	-- with additional identifiability restrictions imposed to accommodate scenarios when only data on response categories are available.]{
		\includegraphics[width=0.645\linewidth]{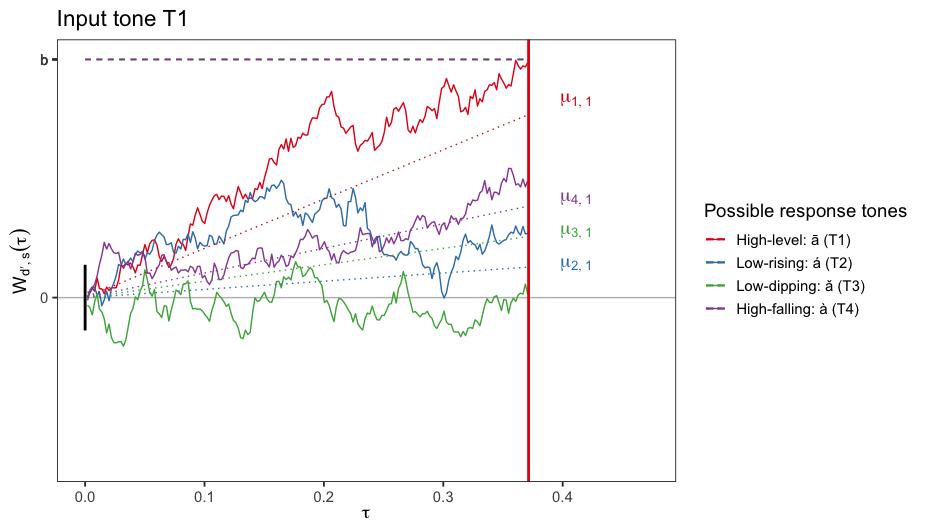}
	}
	\caption{\baselineskip=10pt\small Drift-diffusion model for tone learning.
		The tones \{T1, T2, T3, T4\} represent the different categories; 
		$s$ denotes an input category, $d'$ the different possible response categories, and $d$ the final response category. 
		%A final decision $d$ is arrived when the process $W_{d,s}(\tau)$ crosses its boundary first.} 
		Here we are illustrating a single trial with input tone T1 ($s = 1$) that was eventually correctly identified ($d=1$). 
		Panel (a) shows a process whose parameters can be inferred from data on both response categories and response times.
		Here, after an initial $\delta_{s}$ amount of time required to encode an input category $s$ (here T1),
		the evidence in favor of different possible response categories $d'$ %(here tones T1, T2, T3, T4) 
		accumulates according to latent Wiener diffusion processes $W_{d',s}(\tau)$ (red, blue, green, and purple) with drifts $\mu_{d',s}$. 
		The decision $d$ (here T1) is eventually taken if the underlying process (here the red one) is the first to reach its decision boundary $b_{d,s}$.
		Panel (b) shows a process with additional identifiability restrictions  
		(for all $d'$ and $s$, $\delta_{s}=0$, $b_{d',s}=b$ fixed, and $\sum_{d'=1}^{d_{0}}\mu_{d',s}=d_{0}$) 
		considered in this article 
		which can be inferred from data on response categories alone.
	}
	\label{fig: drift diffusion}
\end{figure}

\paragraph{Drift-diffusion models.} 
A biologically interpretable joint model for decision response accuracies {and associated response times} is 
obtained by imitating the underlying evidence accumulation mechanisms using {latent} drift-diffusion processes racing toward their respective boundaries, 
the process reaching its boundary first producing the final observed decision and the time taken to reach this boundary giving the associated response time (Figure \ref{fig: drift diffusion}, Panel (a)) \citep{usher2001time}. 
%The drift and the boundary parameters then jointly explain the dynamics of decision-making, including the speed-accuracy trade-off (quick vs more accurate decisions, etc.). 

The literature on drift-diffusion processes for decision-making is rather vast 
but is mostly focused on simple binary decision scenarios 
with a single latent diffusion process with two boundaries, 
one for each of the two decision alternatives 
\citep{ratcliff1978theory, ratcliff2016diffusion, smith1988accumulator, ratcliff1998modeling, ratcliff2008diffusion}. 
Multi-category drift-diffusion models with multiple latent processes 
%separate latent processes, one for each decision category, 
are mathematically more easily tractable 
\citep{usher2001time, brown2008simplest, leite2010modeling, dufau2012say, kim2017bayesian} 
but the literature is sparse and focused only on simple static designs. 

Learning to make categorization decisions is, however, a dynamic process, driven by perceptual adjustments in our brain and behavior over time. 
Category learning is thus often studied in longitudinal experiments. %\citep{reetzke2018tracing}. 
To address the need for sophisticated statistical methods for such settings, 
\cite{paulon2021bayes} developed an inverse Gaussian distribution-based multi-category longitudinal drift-diffusion mixed model.

\paragraph{Data requirements and related challenges.} 
Crucially, measurements on both the final decision categories and the associated response times are needed to estimate the drift and the boundary parameters from conventional drift-diffusion models, including the work by \cite{paulon2021bayes}. 
Unfortunately, however,
researchers often only record the participants' decision responses as their go-to measure of categorization performance, 
ignoring the response times \citep{chandrasekaran2014dual,filoteo2010removing}. 
Additionally, eliciting accurate response times can be methodologically challenging, e.g., in the case of experiments conducted online, especially during the Covid-19 pandemic \citep{roark2021auditory}, 
or when the response times from participants/patients are unreliable due to motor deficits \citep{ashby2003category}. 
Participants may also be asked to delay the reporting of their decisions 
so that delayed physiological responses that relate to decision-making can be accurately measured \citep{mchaney2021working}. 
In such cases, the reported response times may not accurately relate to the actual decision times and hence cannot be used in the analysis.  
As a result, conventional drift-diffusion analysis that requires data on both response accuracies and response times, such as \mbox{\cite{paulon2021bayes}}, cannot be used in such scenarios.

\paragraph{The research question.} 
The main research question addressed in this article is to see if a new class of drift-diffusion models can be designed for such scenarios which will allow the biologically interpretable drift-diffusion process parameters to be meaningfully recovered from data on input-output category combinations alone.

%One such data set, described below, motivated the research being reported here. 

%Therefore, conventional drift-diffusion analysis that requires data on response times, such as the one presented in \mbox{\cite{paulon2021bayes}}, cannot be directly applied here. 
%The focus of this article is to see if the drift-diffusion parameters can still be meaningfully recovered from input-output tone categories alone in the PTC1 data.

\paragraph{The inverse-probit model.}
Categorical probability models that build on latent drift-diffusion processes 
can be useful in providing biologically interpretable inference in data sets comprising input-output categories but no response times.
To our knowledge, however, the problem has never been considered in the literature before. 
We aim to address this remarkable gap in this article.  

By integrating out the latent response times from the joint inverse Gaussian drift-diffusion model for response categories and associated response times in \cite{paulon2021bayes}, 
we can arrive at a natural albeit overparametrized model for the response categories.
We refer to this as the `inverse-probit' categorical probability model. 
This inverse-probit model serves as the starting point for the methodology presented in this article 
but, as we describe below, 
it also comes with significant and unique statistical challenges not encountered in the original drift-diffusion model.

\paragraph{Statistical challenges.} 
While scientifically desirable, unfortunately, it is also mathematically impossible to 
infer both the drifts and the boundaries in the inverse-probit model from data only on the decision accuracies.  
We must thus have to keep the values of either the drifts or the boundaries fixed and focus on inferring the other.

However, even when we fix either the drift or the decision boundaries, 
the problem of overparametrization persists. 
In the absence of response times,
only the information on relative frequencies, 
that is empirical probabilities of taking a decision is available.
As the total probability of observing any of the competing decisions is one, 
the identifiability problem remains for the chosen main parameters of interest, 
and appropriate remedial constraints need to be imposed.

Setting an arbitrarily chosen category as the reference provides a simple solution widely adopted in categorical probability models  
but comes with serious limitations, 
including breaking the symmetry of the problem, 
potentially making posterior inference sensitive to the specific choice of the reference category \citep{burgette2012trace,johndrow2013diagonal}.

By breaking the symmetry of the problem, a reference category 
also additionally makes it difficult to infer the potential clustering of the model parameters, especially across different panels. 
To see this, consider a problem with $d_{0}$ categories, with a logistic model for the probabilities $p_{s,d^{\prime}}=\hbox{logistic}(\beta_{s,d^{\prime}}), ~s,d^{\prime}\in\{1:d_{0}\}$, of choosing the ${d^{\prime}}\th$ output category for the $s\th$ input category. 
For each input category $s$, by setting the $s\th$ output category as a reference, 
e.g., by fixing $\beta_{s,s} = 0$, one can then cluster the probabilities of incorrect decision choices, $p_{s,d^{\prime}}$, $d^{\prime}\neq s$. 
However, it is not clear how to compare the probabilities across different input categories (i.e., across the four panels in Figure \ref{fig: accuracies PTC1}), e.g., how to test the equality of $p_{1,1}$ and $p_{2,2}$.

Finally, while coming up with solutions for the aforementioned issues, 
we must also take into consideration the complex longitudinal design of the experiments generating the data. 
Whatever strategy we devise, it should be amenable to a longitudinal mixed model analysis that ideally allows us to 
(a) estimate the smoothly varying longitudinal trajectories of the parameters as the participants learn over time, 
(b) accommodate participant heterogeneity, 
and (c) compare the estimates at different time points 
within and between different input categories.

\paragraph{Our proposed approach.} As a first step toward addressing the identifiability issues and related modeling challenges, 
we keep the boundaries fixed but leave the drift parameters unconstrained.
The decision to focus on the drifts is informed by 
the existing literature on such models cited above where 
the drifts have almost always been allowed more flexibility. 
The analysis of \cite{paulon2021bayes} also showed that 
it is primarily the variations in the drift trajectories that 
explain learning while the boundaries remain relatively stable over time.  
%the heterogeneity between input-response category combinations as well as the heterogeneity between participants. 

As a next step toward establishing identifiability, we apply a `sum to a constant' condition on the drifts so that 
symmetry is maintained in the constrained model.

Implementation of this restriction brings in significant challenges. 
One possibility is to design a prior on the constraint space, a challenging task in itself. 
Additionally, posterior computation for such priors would also be extremely complicated in drift-diffusion models. 
Instead, we conduct inference with an unconstrained prior on the drift parameters and project the samples drawn from the corresponding posterior to the constrained space through a minimal distance mapping.

To adapt this categorical probability model to a longitudinal mixed model setting, 
we then assume 
that the drift parameters comprise input-response-category-specific fixed effects and subject-specific random effects, 
modeling them flexibly by mixtures of locally supported B-spline bases \citep{de1978practical, eilers1996flexible} 
spanning the length of the longitudinal experiment. 
These effects are thus allowed to evolve flexibly as smooth functions of time \citep{ramsay2007applied, morris2015functional, wang2016functional} 
as the participants get more experience and training in their assigned decision tasks.

We take a Bayesian route to estimation and inference. 
Carefully exploiting conditional prior-posterior conjugacy as well as our latent variable construction, 
we design an efficient Markov chain Monte Carlo (MCMC) based algorithm for approximating the posterior, 
where sampling the latent response times for each observed response category greatly simplifies the computations.

We evaluate the numerical performance of the proposed approach in extensive simulation studies. 
We then apply our method to the PTC1 data set described below. 
These applications illustrate the utility of our method 
in providing insights into how the drift parameters characterize the rates of accumulation of evidence 
in the brain evolve over time, 
differ between input-output category combinations, 
as well as between individuals.

\paragraph{Differences from previous works.} 
This article differs in many fundamental ways from all existing works on drift-diffusion models, 
including \mbox{\cite{paulon2021bayes}}, 
where response categories and response times were \emph{both} observed 
and therefore the drift and boundary parameters could be modeled jointly with no identifiability issues. 
In contrast, the current work is motivated by scenarios where data on \emph{only} response categories are available, 
%but no data on response times, 
leading us to the inverse-probit categorical probability model 
which, 
with its complex identifiability issues, 
brings in new unique challenges to performing statistical inference, 
confining us %, unlike \mbox{\cite{paulon2021bayes}}, 
only to infer the drift parameters  
on a relative scale, 
achieved via a novel projection-based approach. 
The introduction and analysis of the inverse-probit model, 
addressing the significant new statistical challenges posed by it, 
ranging across (a) identifiability issues, 
(b) assessment of intra and inter-panel similarities, 
(c) extension to complex longitudinal mixed effects settings 
to accommodate the motivating applications, 
(d) computational implementation of these new models, etc. 
are the novel contributions of this article.

\paragraph{Outline of the article.} 
Section \ref{sec: ptc1} describes our motivating tone learning study. 
Sections \ref{sec: ipm} and \ref{sec: lipmm} develop our longitudinal inverse-probit mixed model.
Section \ref{sec: post inference main} outlines our computational strategies. 
Section \ref{sec: sim studies} presents the results of simulation experiments.
Section \ref{sec: application} presents the results of the proposed method applied to our motivating PTC1 study. 
Section \ref{sec: discussion} concludes the main article with a discussion.
Additional details, including Markov chain Monte Carlo (MCMC) based posterior inference algorithms, are deferred to the supplementary material.

%\vspace*{-20pt}
\section{The PTC1 Data Set} \label{sec: ptc1}
\begin{figure}[!ht]
	\begin{center}
		\includegraphics[width=0.75\linewidth]{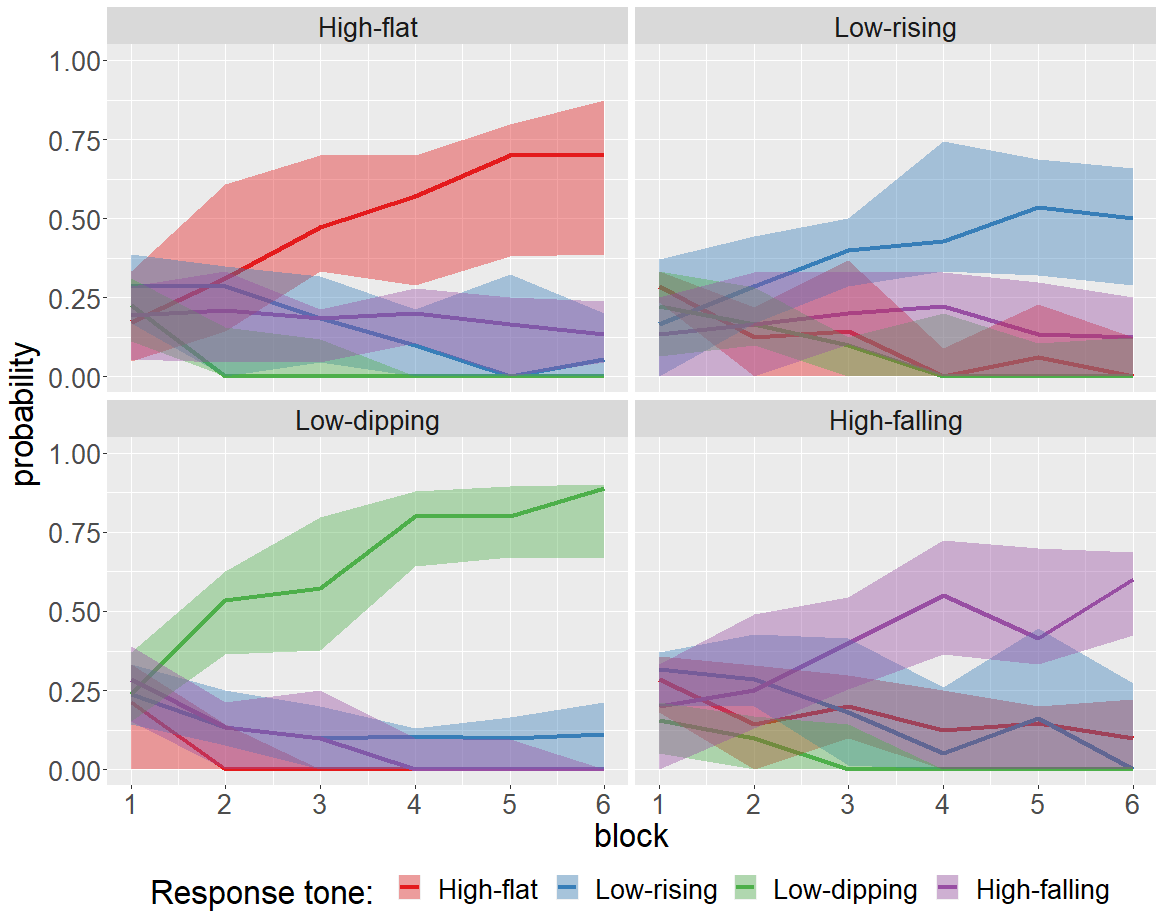} \hspace{0.5cm}
	\end{center}
	\vskip -15pt
	\caption{\baselineskip=10pt\small
		Description of PTC1 data: The proportion of times the response to an input tone was classified into different tone categories over blocks across different subjects, each for the four input tones (indicated in the panel headers).
		The thick line represents the median performance and the shaded region indicates the corresponding middle $30\%$ quantiles across subjects.
	}
	\label{fig: accuracies PTC1} 
\end{figure}

The PTC1 (pupillometry tone categorization experiment 1) data set is obtained from a {Mandarin tone learning} study conducted at the Department of Communication Science and Disorders, University of Pittsburgh \citep{mchaney2021working}. 
Mandarin Chinese is a tonal language, which means that pitch patterns at the syllable level differentiate word meanings. There are four linguistically relevant pitch patterns in Mandarin that make up the four Mandarin tones: high-flat (Tone 1), low-rising (Tone 2), low-dipping (Tone 3), and high-falling (Tone 4). 
For example, the syllable /ma/ can be pronounced using the four different pitch patterns of the four tones, which would result in four different word meanings. 
Adult native English speakers typically experience difficulty differentiating between the four Mandarin tones because pitch contrasts at the syllable level are not linguistically relevant to word meanings in English \citep{wang1999training, wang2003acoustic}. 
Thus, Mandarin tones are valid stimuli to examine how non-native speech sounds are acquired, which has implications for second language learning in adulthood.
In PTC1, a group of native English-speaking younger adults learned to categorize monosyllabic Mandarin tones in a training task. 
During a single trial of training, an input tone was presented over headphones, and the participants were instructed to categorize the tone into one of the four tone categories via a button press on a keyboard. 
Corrective feedback in the form of ``Correct" or ``Wrong" was then provided on screen.
A total of $n=$ 28 participants completed the training task across $T=6$ blocks of training, each block comprising $L=40$ trials. 
Figure \ref{fig: accuracies PTC1} shows the middle $30\%$ quantiles of the proportion of times the response to an input tone was classified into different tone categories over blocks across different subjects, each for the four input tones. 

Pupillometry measurements were also taken during each trial. 
It is commonly used as a metric of 
cognitive effort during listening because increases in pupil diameter are associated with greater usage of cognitive resources \citep{zekveld2011cognitive,peelle2018listening,winn2018best,robison2019pupillometry,parthasarathy2020bottom}. 
One issue with pupillary responses however is that they unfold slowly over time.
In view of that, unlike standard Mandarin tone training tasks, 
where the participants hear the input tone, press the keyboard response, and are provided feedback all within a few seconds \citep{chandrasekaran2016effect, reetzke2018tracing, llanos2020non, smayda2015enhanced}, 
in the PTC1 experiment, there was an intentional four-second delay from the start of the input tone to the response prompt screen where participants made their category decision via button press. 
This four-second delay allows the pupil to dilate in response to hearing the tone and begin to return to baseline before the participant makes a motor response to the button press. 
During this four-second period, participants have likely already made conscious category decisions.  
As such, the response times that are recorded in the end are not meaningful measures of their actual decision times. 

This presents a critical limitation for using these response times for further analysis.
Conventional drift-diffusion analysis that requires data on response times, such as the one presented in \mbox{\cite{paulon2021bayes}}, can no longer be directly applied here. 
The focus of this article is to see if the drift-diffusion parameters can still be meaningfully recovered from input-output tone categories alone in the PTC1 data.

We found drift-diffusion analysis in the absence of reliable data on response times 
challenging enough to merit its separate treatment presented here. 
%The analysis we present in this paper, therefore, excludes the pupillometry measurements. 
Relating drift-diffusion parameters to measures of cognitive effort such as pupillometry is another challenging problem 
that we are pursuing separately elsewhere. 

\section{Inverse-Probit Model} \label{sec: ipm}
The starting point for the proposed inverse-probit categorical probability model follows straightforwardly by integrating out the (unobserved) response times 
from the 
joint model for response categories and associated response times developed in \mbox{\cite{paulon2021bayes}}. 
%The resulting marginal model however requires careful adjustments to identify the model parameters.
The derivation of this original joint model illustrates its latent drift-diffusion process-based underpinnings 
(Figure \ref{fig: drift diffusion}, panel a). 
Later such construction will also be crucial in understanding the diffusion process-based foundations of the marginal categorical probability model 
modified with identifiability constraints proposed in this article 
(Figure \ref{fig: drift diffusion}, panel b). 
We therefore present the derivation from \mbox{\cite{paulon2021bayes}} ditto here 
which also keeps the main paper self-contained.

To begin with, a Wiener diffusion process $W(\tau)$ over domain $\tau \in (0,\infty)$ can be specified as 
$W(\tau) = \mu \tau + \sigma B(\tau)$,
where $B(\tau)$ is the standard Brownian motion, $\mu$ is the drift rate, and $\sigma$ is the diffusion coefficient \citep{cox1965theory, ross1996stochastic}. 
The process has independent normally distributed increments, i.e., $\Delta W(\tau) = \{W(\tau+\Delta\tau) - W(\tau)\} \sim \Normal(\mu \Delta\tau,\sigma^{2} \Delta\tau)$, independently from $W(\tau)$. 
The first passage time of crossing a threshold $b$, $\tau = \inf \{\tau^{\prime}: W(0)=0, W(\tau^{\prime}) \geq b\}$, 
is then distributed according to an inverse Gaussian distribution 
\citep{whitmore1987heuristic, chhikara1988inverse, lu1995degradation} 
with mean $b/\mu$ and variance $b\sigma^2/\mu^{3}$.

Given a perceptual stimulus $s$ and a set of decision choices $d^{\prime}\in\{1:d_{0}\}$, the neurons in the brain accumulate evidence in favor of the different alternatives. 
Modeling this behavior using latent Wiener processes $W_{d',s}(\tau)$ with unit variances, 
assuming that a decision $d$ is made when the decision threshold $b_{d,s}$ for the $d\th$ option is crossed first, 
as illustrated in Figure \ref{fig: drift diffusion}, Panel (a), 
a probability model for the time $\tau_{d}$ to reach decision $d$ 
%for the $d\th$ decision category under the influence of the $s\th$ stimulus 
is obtained as 
%\vspace{-4ex}\\
\be
\textstyle f(\tau_{d} \mid \delta_{s},\mu_{d,s}, b_{d,s})  =  \displaystyle\frac{b_{d,s}}{ \sqrt{2\pi} } (\tau_{d}-\delta_{s})^{-3/2} \exp\left[- \frac{\{b_{d,s}-\mu_{d,s} (\tau_{d}-\delta_{s})\}^{2}}{2  (\tau_{d}-\delta_{s})} \right], \label{eq: drift-diffusion 1}
\ee
%\vspace{-3ex}\\
where $\mu_{d,s}$ denotes the rate of accumulation of evidence, 
$b_{d,s}$ the decision boundaries, 
and $\delta_{s}$ an offset representing time not directly related to the underlying evidence accumulation processes (e.g., the time required to encode the $s\th$ signal before evidence accumulation begins, 
%the time to press a computer key to record a response after a decision is reached, 
etc.). 
We let $\btheta_{d^{\prime},s}=(\delta_{s},\mu_{d^{\prime},s},b_{d^{\prime},s})\trans$. 

{\bf Joint model for $(d,\tau)$:} 
Since a decision $d$ is reached at response time $\tau$ if the corresponding threshold is crossed first, that is when $\{\tau = \tau_{d}\} \cap_{d^{\prime} \neq d} \{\tau_{d^{\prime}} > \tau_{d}\}$, we have $d = \arg \min_{d^{\prime}\in\{1:d_{0}\}} \tau_{d^{\prime}}$. 
Assuming simultaneous accumulation of evidence for all decision categories, 
modeled by independent Wiener processes, 
and termination when the threshold for the observed decision category $d$ is reached, 
the joint distribution of $(d, \tau)$ is thus given by 
%\vspace{-4ex}\\
\be
& f(d, \tau \mid s,\btheta) = g(\tau \mid \btheta_{d,s}) \prod_{d^{\prime} \neq d} \{1 - G(\tau \mid \btheta_{d^{\prime},s})\},   \label{eq: inv-Gaussian likelihood}
\ee
%\vspace{-4ex}\\
where, to distinguish from the generic notation $f$, we now use $g(\cdot \mid \btheta)$ and $G(\cdot \mid \btheta)$ to denote, 
respectively, the probability density function (pdf) and the cumulative distribution function (cdf) of an inverse Gaussian distribution, as defined in (\ref{eq: drift-diffusion 1}).

{\bf Marginal model for $d$:}  
When the response times $\tau$ are unobserved, the probability of taking decision $d$ given the stimulus $s$ is thus obtained from \eqref{eq: inv-Gaussian likelihood} by integrating out the $\tau$ as 
\vspace{-4ex}\\
\be
& P(d \mid s,\btheta) = \int_{\delta_{s}}^{\infty} g(\tau \mid \btheta_{d,s}) \prod_{d^{\prime} \neq d} \left\{1 - G(\tau \mid \btheta_{d^{\prime},s})\right\} d\tau. \label{eq: multinomial inv-probit}
\ee
\vspace{-4ex}\\
The construction of model (\ref{eq: multinomial inv-probit}) is similar to traditional multinomial probit/logit regression models except that the latent variables are now inverse Gaussian distributed as opposed to being normal or extreme-value distributed, 
and the observed category is associated with the minimum of the latent variables in contrast to being identified with the maximum of the latent variables. 
We thus refer to this model as a `multinomial inverse-probit model'.

With data on both response categories $d$ and response times $\tau$ available, the joint model (\ref{eq: inv-Gaussian likelihood}) was used to construct the likelihood function in \mbox{\cite{paulon2021bayes}}.
In the absence of data on the response times $\tau$, however, 
the inverse-probit model in (\ref{eq: multinomial inv-probit}) provides the basic building block for constructing the likelihood function for the observed response categories. 
As mentioned in the Abstract, discussed in the Introduction, and detailed in Section \ref{sec: stat background} below, the {marginal} inverse-probit model (\ref{eq: multinomial inv-probit}) for observed categories brings in many new identifiability issues and inference challenges not originally encountered for the joint model (\ref{eq: inv-Gaussian likelihood}) developed in \mbox{\cite{paulon2021bayes}}.  
Solving these new challenges for the marginal model (\ref{eq: multinomial inv-probit}) 
to infer the underlying drift-diffusion parameters $\btheta_{d^{\prime},s}$, for all $d^{\prime}$, is the focus of this current article.

\subsection{Identifiability Issues and Related Modeling Challenges} 
\label{sec: stat background}

To begin with, we note that model (\ref{eq: multinomial inv-probit}) in itself cannot be identified from data on only the response categories. 
The offset parameters can easily be seen to not be identifiable since $P\left(\tau_{d}\leq \wedge_{d^{\prime}} \tau_{d^{\prime}}\right)=P\left\{(\tau_{d}-\delta)\leq \wedge_{d^{\prime}} \left(\tau_{d^{\prime}}-\delta\right)\right\}$ for any $\delta$, where $\wedge_{d^{\prime}}\tau_{d^{\prime}}$ denotes the minimum of $\tau_{d^{\prime}}, d^{\prime}\in\{1:d_{0}\}$.  
As is also well known in the literature, in categorical probability models, the location and scale of the latent continuous variables are not also separately identifiable. 
The following lemma establishes these points for the inverse-probit model.

\begin{Lem}\label{lem1}
	The offset parameters $\delta_{s}$ are not identifiable in model (\ref{eq: multinomial inv-probit}).
	The drift and the boundary parameters, respectively $\mu_{d^{\prime},s}$ and $b_{d^{\prime},s}$, are also not separately identifiable in model (\ref{eq: multinomial inv-probit}). 
\end{Lem}

In the proof of Lemma \ref{lem1} given in \ref{app: lem1}, 
we have specifically shown that 
$ P\left(d\mid s, \btheta \right) = P\left(d\mid s, \btheta^{\star} \right),$
where the drift and boundary parameters in $\btheta=\left\{ \left( \mu_{d^{\prime},s}, b_{d^{\prime},s}\right); d^{\prime}=1,\ldots, d_{0}\right\}$ and $\btheta^{\star}=\left\{ \Big( \mu_{d^{\prime},s}^{\star}, b_{d^{\prime},s}^{\star}\Big); d^{\prime}=1,\ldots, d_{0}\right\}$ satisfy 
$\mu_{d^{\prime},s}^{\star}=c \mu_{d^{\prime},s}$ and $b_{d^{\prime},s}^{\star}= c^{-1} b_{d^{\prime},s}$ for some constant $c>0$.
The result follows by noting that 
the transformation $\tau_{d^{\prime},s}^{\star} = {c}^{-2}\tau_{d^{\prime},s}$ 
does not change the ordering between the $\tau_{d^{\prime},s}$'s 
and hence the probabilities of the resulting decisions 
$d = \arg \min_{d^{\prime}\in\{1:d_{0}\}} \tau_{d^{\prime}} = \arg \min_{d^{\prime}\in\{1:d_{0}\}} \tau_{d^{\prime}}^{\star}$ also remain the same. 
This has the simple implication that if the rate of accumulation of evidence is faster, 
then the same decision distribution is obtained if the corresponding boundaries are accordingly closer and conversely.

In fact, given the information on input and output categories alone, 
if $d_{0}$ denotes the number of possible decision categories, at most $d_{0}-1$ parameters are estimable. 
To see this, consider the probabilities $P(d^{\prime} \mid s,\btheta)$, $d^{\prime}=1, \ldots, d_{0}$, 
where $\btheta$ is the $m$-dimensional vector of parameters, 
possibly containing drift parameters and decision boundaries.    
Given the perceptual stimulus $s$ as input, the probabilities satisfy $\sum_{d^{\prime}=1}^{d_{0}}P(d^{\prime} \mid s,\btheta)=1$. 
Thus, the function $P_{s}(\btheta)= \{P(1 \mid s, \btheta), \cdots, P(d_{0}\mid s, \btheta)\}\trans$ lie on a $d_{0}-1$-dimensional simplex,
$P_{s}(\btheta): \btheta \rightarrow \Delta^{d_{0}-1}$, and by the model in (\ref{eq: multinomial inv-probit}) the mapping is continuous. 
Thus, it can be shown by the {\it Invariance of Domain} theorem  \citep[see, e.g.,][]{Deo2018Algebraic} that if $P_{s}$ is injective and continuous, 
then the domain of $P_{s}$ must belong to $\mathbb{R}^{m}$, 
where $m\leq d_{0}-1$. 
Thus in order to ensure identifiablilty of $\left\{ P_{s} (\btheta); \btheta \right\}$, 
we must parametrize the probability vector with at most ${d_{0}-1}$ parameters.

The existing literature on drift-diffusion models discussed in the Introduction has traditionally put more emphasis on modeling the drifts 
(as their reference in the literature as `drift'-diffusion models suggests). 
Previous research on joint models for response tones and associated response times in \cite{paulon2021bayes} also suggest that the boundaries remain stable around a value of $2$
and it is primarily the changes in the drift rates that explain longitudinal learning. 
In view of this, we keep the boundaries fixed at the constant $b=2$ and treat the drifts to be the free parameters instead. 
In our simulations and real data applications, 
it is observed that the estimates of the drift parameters and the associated cluster configurations are not very sensitive to small-to-moderate deviations of $b$ around $2$. 
In our codes implementing our method, available as part of the supplementary materials, we allow the practitioner to choose a value of $b$ as they see fit for their specific application.
The latent drift-diffusion process based with these constraints, namely $\delta_{s} = 0$ and $b_{d^{\prime},s} = b$ for all $d^{\prime}$, 
is shown in Figure \ref{fig: drift diffusion}, Panel (b).

While fixing $\delta_{s}=0$ and $b_{d^{\prime},s}$ to some known constant $b$ reduces the size of the parameter space to $d_{0}$, 
to ensure identifiability, 
we still need at least one more constraint on the drift parameters $\bmu_{1:d_{0},s}$.
In categorical probability models, 
the identifiabily problem of the location parameter is usually addressed by setting one category as a reference and modeling the probabilities of the others \citep{Albert1993Bayesian,chib1998analysis,borooah2002logit,agresti2018introduction}. 
However, posterior predictions from Bayesian categorical probability models with asymmetric constraints may be sensitive to the choice of reference category \citep[see][]{burgette2012trace,johndrow2013diagonal}.
Further, as also discussed in the Introduction,
the goal of clustering the drift parameters $\bmu_{1:d_{0},s}$ across $s$ can not be accomplished by this apparently simple solution.  

The problem can be addressed by imposing a symmetric constraint on $\bmu_{1:d_{0},s}$ instead.
%Symmetric identifiability constraints have been previously proposed by \cite{burgette2010symmetric}. 
%Their {\it sum to zero} constraint solves the problem of relying on a single base category, instead, it is adaptively chosen by the model. 
A symmetric identifiability constraint has been previously proposed by \cite{Burgette2021A} in the context of multinomial probit models, 
where they considered a sum-to-zero constraint on the latent utilities. 
To implement the constraint, they introduced a \emph{faux base category} indicator parameter, 
which is assigned a discrete uniform prior and then learned via MCMC.
Given this {faux base category} indicator, 
the other parameters are adjusted so that the sum-to-zero restriction is satisfied. 
However, the introduction of a base category, even if adaptively chosen, 
does not facilitate the clustering of $\bmu_{1:d_{0},s}$ within and across the different input categories $s$. 

\subsection{Our Proposed Approach} \label{sec: prop approach} 
In coming up with solutions for these challenges, 
we take into consideration the complex design of our motivating tone learning experiments,
so that our approach is easily extendable to longitudinal mixed model settings, 
allowing us to 
(a) estimate the smoothly varying trajectories of the parameters as the participants learn over time, 
(b) accommodate the heterogeneity between the participants, 
and 
(c) compare between the estimates not just within but also crucially between the different panels.

Similar to the {\it sum to zero} constraint in the multinomial probit model of \cite{Burgette2021A}, 
we impose a symmetric {\it sum to a constant} constraint on the drift parameters $\bmu_{1:d_{0},s}$ to identify our new class of inverse-probit models, 
although our implementation is quite different from theirs. 
To conduct inference, we start with an unconstrained prior, 
then sample from the corresponding unconstrained posterior, 
and finally project these samples to the constrained space through a minimal distance mapping. 
Similar ideas have previously been applied %as a post-processing step in order 
to satisfy natural constraints in other contexts. 
See, e.g., \cite{Dunson2003Bayesian, Gunn2005Transformation}. 

This approach is significantly advantageous both from a modeling and a computational perspective. 
On one hand, the basic building blocks are relatively easily extended to complex longitudinal mixed model settings,
on the other, posterior computation is facilitated as this allows 
the use of conjugate priors for the unconstrained parameters.  
Projection of the drift parameters onto the same space further makes them directly comparable, 
allowing clustering within and across the panels. 
The projected drifts can now be interpreted only on a relative scale 
but such compromises are not avoidable given the challenges we face.

\subsubsection{Minimal Distance Mapping} \label{sec: MD Projection} 

As the drift parameters are positive, the {\it sum to a constant $k$} constraint leads to the constrained space $\S_{k}=\{\bmu: {\bf 1}^{T} \bmu=k, ~\mu_{j}>0, ~j=1,\ldots,d_{0}\}$ on which $\bmu_{1:d_{0},s}$ should be projected.
The space $\S_{k}$ is semi-closed, and therefore, the projection of any point $\bmu$ onto $\S_{k}$ may not exist. 
As a simple one dimensional example, let $x=-1$ and $\mathcal{S}=(0,1]$, then $\arg\min_{y\in \mathcal{S}}|y-x|=0\notin \mathcal{S}$. 
Further, from a practical perspective, 
a drift parameter infinitesimally close to zero makes the distribution of the associated response times very flat which is typically not observed in real data.
Therefore, we choose a small $\varepsilon>0$ and project $\bmu$ onto 
$ \S_{\varepsilon, k}=\{\bmu: {\bf 1}^{T} \bmu=k, ~\mu_{j}\geq \varepsilon , ~j=1,\ldots,d_{0}\}.$
We then define the projection of a point $\bmu$ onto $\S_{\varepsilon,k}$ through minimal distance mapping as 
%\vspace{-5ex}\\
\bse 
\bmu^{\star}=\mathrm{Proj}_{ \S_{\varepsilon, k}}(\bmu) :=\left\{ \mathrm{argmin}_{\bnu} \|\bmu-\bnu\|: \bnu\in \S_{\varepsilon,k} \right\},
\ese
%\vspace{-5ex}\\
{where $\|\cdot\|$ is the Euclidean norm.}
Note that for appropriate choices of $(k,\varepsilon)$, $\S_{\varepsilon,k}$ is non-empty, closed and convex. 
Therefore, $\bmu^{\star}$ exists and is unique by the Hilbert projection theorem \citep{Rudin1991Functional}.
The solution to this projection problem comes from the following result from \citet{Beck2017First}. 

\begin{Lem}\label{lem:proj1}
	Let $ \S_{\varepsilon, k}$ be as defined above, and $\S_{\varepsilon}=\left\{\bmu: \mu_{j}\geq \varepsilon,~j=1,\ldots,d_{0} \right\}$. 
	Then, $\mathrm{Proj}_{ \S_{\varepsilon, k}}(\bmu)=\mathrm{Proj}_{ \S_{\varepsilon}}(\bmu - u^{\star} {\bf 1})$, where 
	$u^{\star}$ is a solution to the equation $ {\bf 1}^{T} \mathrm{Proj}_{ \S_{\varepsilon}}(\bmu - u^{\star} {\bf 1})=k $.
\end{Lem}
%A proof of Lemma \ref{lem:proj1} is given in \citet{Beck2017First}.
Although the analytical form of the solution is not available, as is evident from the above result, the solution mainly relies on finding a root $u^{\star}$ of the non-increasing function $\phi(u^{\star})={\bf 1}^{T} \mathrm{Proj}_{ \S_{\varepsilon}}(\bmu - u^{\star} {\bf 1})-k$.
We apply an algorithm {based on} \cite{Duchi2008Efficient} to reach the solution. 
The algorithm is described in \ref{app: algo mindist}.

\subsubsection{Identifiability Restrictions}\label{sec: identifiability}

The projection approach solves the problem of identifiability and maps the probability vector corresponding to an input tone $s$ to the constraint space of $\bmu_{1:d_{0},s}$, $\S_{\varepsilon,k}$. 
The following theorem shows that the mapping from the constrained space of $\bmu_{1:d_{0},s}$ to the probability vector $\bP (\bmu_{1:d_{0},s})=\{p_{1}(\bmu_{1:d_{0},s}), \dots, p_{d_{0}}(\bmu_{1:d_{0},s}) \}\trans$ is injective. 
To keep the ideas simple, we consider the domain of the function to be $\S_{0,k}$ (i.e., $\varepsilon=0$) instead of $\S_{\varepsilon,k}$ although a very similar proof would follow if $\S_{\varepsilon,k}$ were considered.

\begin{Thm}\label{thm:injection}
	Let $p_{d}(\bmu_{1:d_{0},s})$ be the probability of observing the output tone $d$ given the input tone $s$ and the drift parameters $\bmu_{1:d_{0},s}$, as given in (\ref{eq: multinomial inv-probit}), for each $d=1:d_{0}$. 
	Suppose $\bmu_{1:d_{0},s}$ lies on the space $\S_{0,k}$. 
	Then, the function from $\S_{0,k}$ to the space of probabilities $\left\{p_{d}(\bmu_{1:d_{0},s}); d=1:d_{0}\right\}$ is injective.   %\colred{We don't need to change the index $d\to d^{\prime}$ here, because here $d$ indicates the observed output tone.}
\end{Thm}
A proof is presented in \ref{app: thm1}.

\subsubsection{Conjugate Priors for the Unconstrained Drifts}

From (\ref{eq: multinomial inv-probit}), 
given $\tau_{1},\ldots,\tau_{d_{0}}$, such that $\tau_{d}\leq \min \left\{ \tau_{1},\ldots,\tau_{d_{0}} \right\}$, the posterior full conditional of $\bmu_{1:d_{0},s}$ is proportional to 
$\pi_{\bmu}^{(n)} \propto \pi ( \bmu_{1:d_{0},s} ) \times \prod_{d^{\prime}=1}^{d_{0}} g\left(\tau_{d^{\prime}} ~ \mid  \mu_{d^{\prime},s}\right),$ where $\pi(\cdot)$ is the prior of $\bmu_{1:d_{0},s}$.
Observe that $\prod_{d^{\prime}=1}^{d_{0}} g\left(\tau_{d^{\prime}} ~ \mid  \mu_{d^{\prime},s}\right)$ is Gaussian in $\bmu_{1:d_{0},s}$.
A Gaussian prior on $\bmu_{1:d_{0},s}$ thus 
induces a conditional posterior for $\bmu_{1:d_{0},s}$ that is also Gaussian and hence very easy to sample from.
Importantly, these benefits also extend naturally to multivariate Gaussian priors for any parameter vector $\bbeta_{d^{\prime},s}$ that relates to $\mu_{d^{\prime},s}$ linearly. 
This will be crucial in allowing us to extend the basic building block to longitudinal functional mixed model settings in Section \ref{sec: lipmm} next, 
where we will be modeling time varying $\mu_{d^{\prime},s}(t)$ 
as flexible mixtures of B-splines with associated coefficients $\bbeta_{d^{\prime},s}$.

\subsubsection{Justification as a Proper Bayesian Procedure} 
Define the constrained conditional posterior distribution, $\tilde{\pi}_{\tilde{\bmu}}^{(n)}$, of the drift parameters $\bmu$ as
\vspace{-4ex}\\
\bse
\tilde{\pi}_{\tilde{\bmu}}^{(n)} \left(B \mid \bzeta \right)= \pi_{\bmu}^{(n)} \left(\left\{ \bmu: \mathrm{Proj}( \bmu )\in B \right\} \mid \bzeta \right), \quad B \subseteq \S_{\varepsilon,k},
\ese
\vspace{-4ex}\\
\noindent where $\pi_{\bmu}^{(n)}$ is the unconstrained conditional posterior of $\bmu_{1:d_{0},s}$, given the other variables $\bzeta$.
The analytic form of the constrained conditional posterior is not available.

\cite{sen2018constrained} established a proper Bayesian justification for the posterior projection approach by showing the existence of a prior $\tilde{\pi}\left(\bmu_{1:d_{0},s}\right)$ on the constrained space $\S_{\varepsilon,k}$ 
such that the resulting posterior is the same as the projected posterior $\tilde{\pi}_{\tilde{\bmu}}^{(n)}$. 
When $\S_{\varepsilon,k}$ is non-empty, closed, and convex, i.e., the projection operator is measurable, such a prior exists if the unconstrained posterior is absolutely continuous with respect to the unconstrained prior \citep[][Corollary 1]{sen2018constrained}. 
As the unconstrained induced prior and posterior of the drift parameters are both Gaussian, this result holds in our case as well.

\section{Extension to Longitudinal Mixed Models} \label{sec: lipmm}

In this section we adapt the inverse probit model discussed in Section \ref{sec: ipm} to complex longitudinal design of our motivating PTC1 data set described in the Introduction.
Let $s_{i,\ell,t}$ denote the input tone for the $i\th$ individual in the $\ell\th$ trial of block $t$. 
Likewise, let $d_{i,\ell,t}$ denote the output tone selected 
by the $i\th$ individual in the $\ell\th$ trial of block $t$. 
Setting the offsets at zero, and boundary parameters to a fixed constant $b$, 
we now have 
%\vspace{-4ex}
\be
%\begin{split} 
P\{d_{i,\ell,t} =d \mid s_{i,\ell,t}=s, \bmu_{1:d_{0},s}^{(i)}(t)\} = \int_{0}^{\infty} g\{\tau \mid \mu_{d,s}^{(i)}(t)\} \prod_{d^{\prime} \neq d} \left[1 - G\{\tau \mid \mu_{d^{\prime},s}^{(i)}(t)\}\right] d\tau, \label{eq: mod1b}\\
\mbox{where}~ ~ g\{\tau \mid s_{i,\ell,t}=s, \mu_{d^{\prime},s}^{(i)}(t)=\mu\}  \textstyle = \displaystyle\frac{b}{\sqrt{2\pi} \tau^{3/2}}  \exp\left[- \frac{\{b-\mu \tau\}^{2}}   {2 \tau} \right].  \hspace{1 in}\notag
%\end{split} 
\ee
\vspace{-4ex}\\
The drift rates $\mu_{d^{\prime},s}^{(i)}(t)$ now vary with the blocks $t$. 
In addition, we accommodate random effects by allowing $\mu_{d^{\prime},s}^{(i)}(t)$ to also depend on the subject index $i$. 
We let 
$\bd=\{d_{i,\ell,t}\}_{i,\ell,t}$, and $d_{0}$ be the number of possible decision categories (T1, T2, $\cdots$, T${d_{0}}$). 
The likelihood function 
thus takes the form  
\vspace{-4ex}\\
\bse
L(\bd \mid \bs, \btheta) = \prod_{d=1}^{d_{0}}  \prod_{s=1}^{d_{0}} \prod_{t=1}^{T}\prod_{i=1}^{n} \prod_{\ell=1}^{L} \left[ P\{d_{i,\ell,t} \mid s_{i,\ell,t}, \bmu_{1:d_{0},s}^{(i)}(t)\} \right]^{1\{d_{i,\ell,t} = d, s_{i,\ell,t}=s\}}. 
\ese

We reiterate that 
in deriving the identifiability conditions and designing their implementation strategy in Section \ref{sec: prop approach}, 
we had to make sure that they would be applicable to the complex multi-subject longitudinal design of the PTC1 data set.
Following those ideas, 
we model the time-varying mixed effects drift parameters $\mu_{d^{\prime},s}^{(i)}(t)$ without any constraints first, 
then project them to the space satisfying the necessary identifying conditions.

For the unconstrained model, 
we follow the outline of \cite{paulon2021bayes} 
with necessary likelihood adjustments. 
The details are deferred to Section S.1 of the supplementary material. 
We present here a general outline.

We decompose $\mu_{d^{\prime},s}^{(i)}(t)  =  f_{d^{\prime},s}(t) + u_{d^{\prime},s}^{(i)}(t)$
where $f_{d^{\prime},s}(t)$ and $u_{d^{\prime},s}^{(i)}(t)$ denote, respectively, 
fixed and random effects components, 
which are both modeled using flexible mixtures of B-spline bases.  
This allows us to cluster the fixed effects for different $(d^{\prime},s)$ combinations with similar shapes by clustering the corresponding B-spline coefficients.

Given posterior samples of $f_{d^{\prime},s}(t)$ and $u_{d^{\prime},s}^{(i)}(t)$,  unconstrained samples of $\mu_{d^{\prime},s}^{(i)}(t)$ are obtained. 
For every input tone $s$, these unconstrained $\bmu_{1:d_{0},s}^{(i)}(t)$'s are then projected to the space $\S_{\varepsilon,k}$ following the method described in Section \ref{sec: MD Projection}.

\section{Posterior Inference} \label{sec: post inference main}

Posterior inference for our proposed inverse-probit mixed model is carried out using samples drawn from the posterior using MCMC algorithm.  
The algorithm carefully exploits the conditional independence relationships encoded in the model 
as well as the latent variable construction of the model.

Inference can be greatly simplified by sampling the passage times $\tau_{1:d_{0}}$ and then conditioning on them. 
However, 
it is not possible to generate $\tau_{1:d_{0}}$ sequentially, 
e.g., by generating the passage time of the $d$-th decision choice $\tau_{d}$ independently, 
and that of the other decision choices from a truncated inverse-Gaussian distribution, 
left truncated at $\tau_{d}$.\footnote{We can see this in a simpler example. Suppose we are interested in generating a sample from the conditional distribution of $\btau=(\tau_{1},\tau_{2})$ given $d=\arg\min_{j} \tau_{j}=1$, where $\tau_{i}~{\sim}~\mathtt{Uniform}(0,1)$, $i=1,2$, independently. The conditional density of $\boldsymbol{\tau}$ given $d=1$ is $f_{\boldsymbol{\tau}\mid d} (\tau_{1},\tau_{2})= 2$ if $0<\tau_{1}\leq \tau_{2}<1$, and $=0$ otherwise. However, if we draw $\tau_{1}$ from $\mathtt{Uniform}(0,1)$ first and let that realization be $\tau^{\star}$, and draw $\tau_{2}$ from the truncated uniform distribution (left truncated at $\tau^{\star}$), then the pdf of the realization of $(\tau_{1},\tau_{2})$ is $\tau^{\star-1}$.}

We implement a simple accept-reject sampler instead 
which generates values from the joint distribution of $\tau_{1:d_{0}}$ and accepts the sample if $\tau_{d}\leq \tau_{1:d_{0}}$. 
It is fast and produces a sample from the desired target conditional distribution. 
We formalize this result in the following lemma. 

\begin{Lem}
	\label{lem3}
	Let $g\left(\tau_{1:d_{0}}\mid \mu_{1:d_{0}} \right)$ be the joint distribution of $\tau_{1:d_{0}}$.
	Consider the following accept-reject algorithm: 
	\begin{algorithm}[H] 
		\caption{Generating the passage times $\tau_{1:d_{0}}$ given $\arg\min_{d^{\prime}\in\{1:d_{0}\}}\tau_{d^{\prime}}=d$}
		\label{algo: RT}
		\begin{algorithmic}[1]
			\vspace{0.2cm}
			\State Generate $\tau_{1:d_{0}}$ from the joint distribution $g\left(\tau_{1:d_{0}}| \mu_{1:d_{0}} \right)$.
			\State Accept $\tau_{1:d_{0}}$ if $\tau_{d}\leq \tau_{1:d_{0}}$.
			\State Return to Step 1 otherwise.
		\end{algorithmic} 
	\end{algorithm}
	
	\noindent Algorithm \ref{algo: RT} generates samples from the conditional joint distribution of $\tau_{1:d_{0}}$, conditioned on the event $\tau_{d}\leq \tau_{1:d_{0}}$.
\end{Lem}

Proof of Lemma \ref{lem3} is provided in \ref{app: lem3}. 

It can be verified that the acceptance ratio of Algorithm \ref{algo: RT} is $M^{-1}=P\left( \tau_{d} \leq \tau_{1:d_{0}}\right)$ \citep[see][]{Robert2004Monte} 
which depends on the drift parameters alone. 
If the drift parameters are ordered accordingly, so as to satisfy $\mu_{d} \geq \mu_{1:d_{0}}$, the acceptance ratios increase. 
The algorithm thus becomes faster as the sampler converges.  

As noted earlier, sampling the latent inverse-gaussian distributed response times $\tau_{1:d_{0}}$ greatly simplifies computation. 
Most of the chosen priors, including the priors on the coefficients $\bbeta$ in the fixed and random effects, are conjugate.
Due to space constraints, the details are deferred to Section S.3 in the supplementary material.

\section{Simulation Studies} \label{sec: sim studies}

In this section, we discuss the results of a synthetic numerical experiment. 
We simulate data from a complex longitudinal design that mimics the real PTC1 data set. 
Our generating model contains fixed effects components attributed to different input-response tone combinations and random components attributed to individuals.

We recall that our main objective here is to identify the similarities and differences between the underlying brain mechanisms associated with different input-response category combinations over time 
while also assessing their individual heterogeneity, 
as characterized by latent drift-diffusion processes 
whose parameters can be biologically interpreted. 
The estimation of the probability curves for different input-response combinations, 
while a good indicator of our model's fit, 
is not the main purpose of this endeavor. 
Traditional categorical probability models, 
such as multinomial probit or logit, 
are thus not relevant to the scientific problem we are trying to address here. 
We are also not aware of any other work in the drift-diffusion literature 
that attempts to estimate the underlying parameters from category response data alone. 
In view of this, we restrict our focus to evaluating the performance of the proposed 
biologically meaningful longitudinal inverse-probit mixed model 
but do not present comparisons with any other model.

\paragraph{Design.} 
In designing the simulation scenario, we have tried to mimic our motivating category learning data sets. 
We chose $n=20$ as the number of participants being trained over $T=10$ blocks to identify $d_{0}=4$ tones. For each input tone and each block, there are $L=40$ trials.
We set the true $\mu_{d^{\prime},s}(t)$ values in such a way that they are far from satisfying the constraint $\sum_{d^{\prime} =1:d_{0}}\mu_{d^{\prime},s}=k$, and the decision boundary is set to $b=2$ for all $(d^{\prime},s)$.
The true drift parameters and the true probabilities, averaged over the participants of each input-response category combination are shown in Figure \ref{fig: simulated_data_description}.

There are four true clusters in total, two for correct categorizations, {\small$S_{1}, ~S_{2}$}, and two for incorrect categorizations, {\small$M_{1}, ~M_{2}$},
as follows:
{\small $S_{1}=\{(1,1),(2,2)\}$, $S_{2}=\{(3,3),(4,4)\}$, $M_{1}=\{(1,2),(1,3),(2,1),(2,3),(3,4),(4,3)\}$, $M_{2}=\{(1,4),(2,4),(3,1),(3,2),$ $(4,1),(4,2)\}$}. 
We may interpret $M_{1}$ as the cluster of difficult alternatives, and $M_{2}$ as the cluster of easy alternatives.
Thus, there are similarities in overall trajectories of $\{T_{1},T_{2}\}$ and $\{T_{3},T_{4}\}$, 
differentiating between easy and hard category recognition problems.  
We experimented with $50$ synthetic data sets generated according to this design.

\afterpage{
	\begin{figure}[ht!]
		\centering
		\hskip 0.5cm\includegraphics[scale=.35, trim=2cm 0.25cm 1cm 0.25cm]{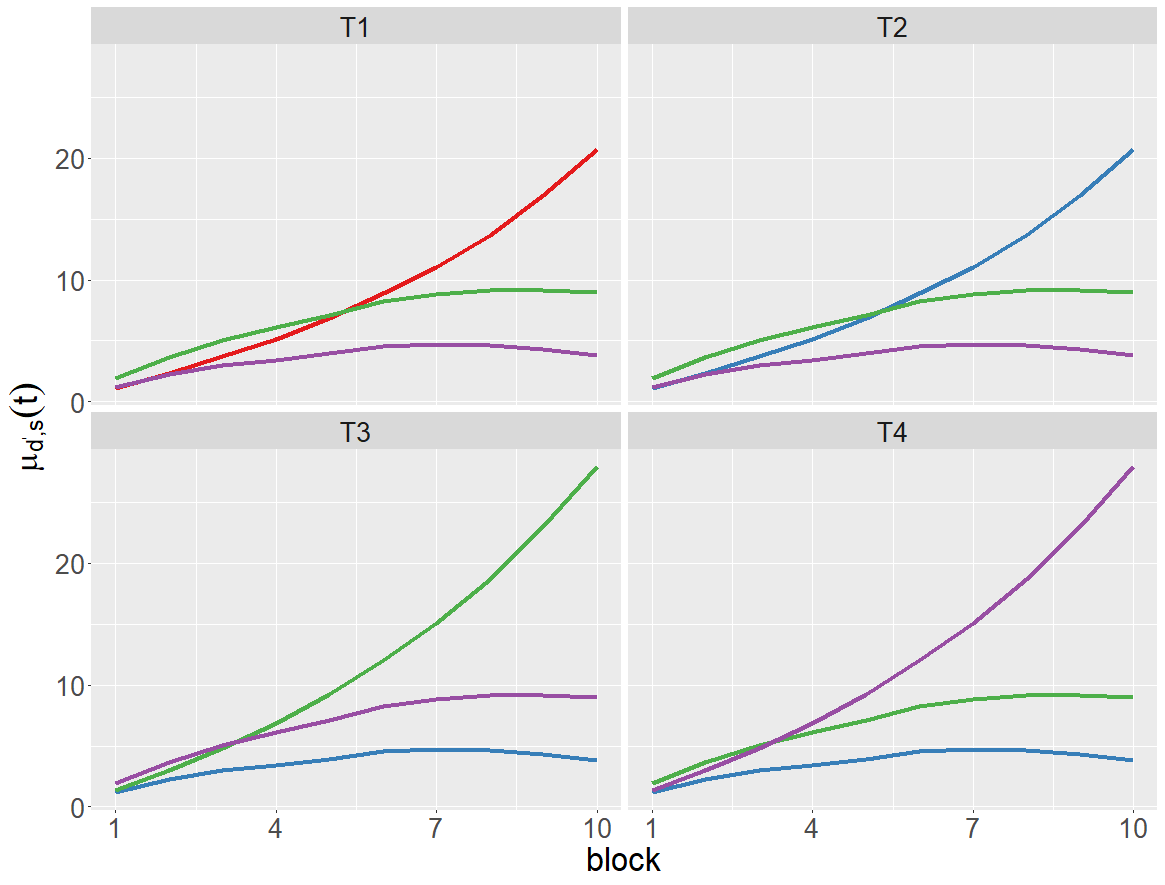} \\\vskip 0.5cm
		\hskip 0.5cm\includegraphics[scale=.355, trim=2cm 0.25cm 1cm 0.25cm]{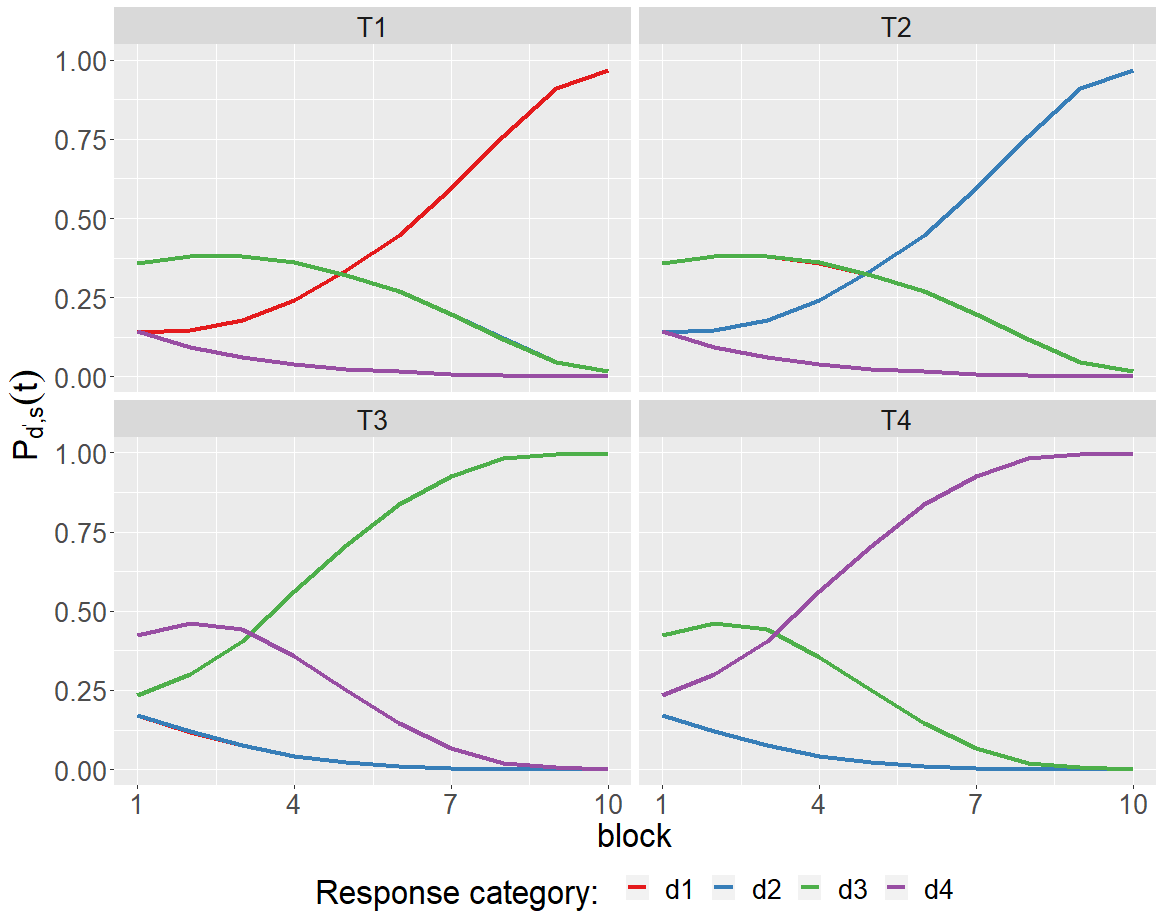}
		\caption{\baselineskip=10pt\small 
			Description of the synthetic data: 
			True values of the drift parameters averaged over the subjects, denoted by $\mu_{d^{\prime},s}(t)$, 
			and true probabilities $P\{d_{i,\ell,t} \mid s_{i,\ell,t}, \bmu_{1:d_{0},s}^{(i)}(t)\}$ averaged over the subjects, denoted here by $P_{d^{\prime},s}(t)$.
			Here T1, T2, T3, and T4 represent input categories 1 to 4, respectively. 
			Some of the curves overlap according to the true clustering structure described in Section \ref{sec: sim studies}. 
		}
		\label{fig: simulated_data_description}
	\end{figure}
}

\afterpage{
	\begin{figure}[ht!]
		\centering
		\hskip 0.5cm\includegraphics[width=0.75\textwidth, trim=2cm 0.25cm 1cm 0.25cm]{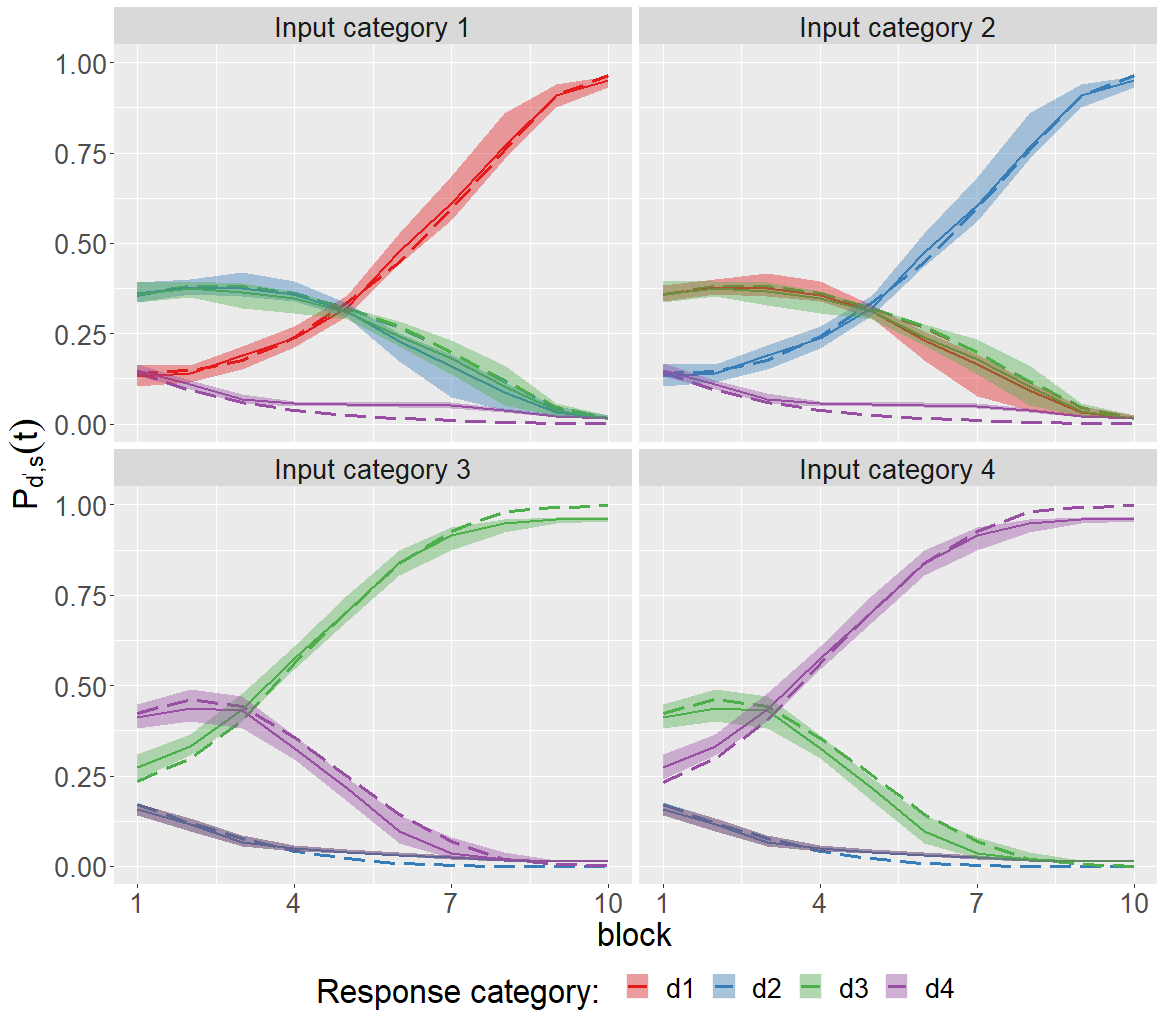}
		\caption{\baselineskip=10pt\small 
			Results for synthetic data: 
			Posterior trajectories of the probabilities for each combination of $(d^{\prime},s)$ over blocks estimated by the proposed model. 
			The shaded areas represent the corresponding $95\%$ point-wise credible intervals. 
			The thick dashed lines represent underlying true curves 
			some of which overlap according to the true clustering structure described in Section \ref{sec: sim studies}. 
		}
		\label{fig: simulation_estimated_prob}
	\end{figure}
}
\paragraph{Results.} 
As the true drift parameters themselves do not satisfy the constraint, and the estimated drift parameters are on the constrained space, we cannot validate our method by its predictive performance of the drift parameters. 
Instead, the proposed method is validated in terms of the estimated probabilities.

Figure \ref{fig: simulation_estimated_prob} 
shows the estimated posterior probability trajectories along with the $95\%$ credible interval and the underlying true probability curves for every combination $(d^{\prime},s)$ in a typical scenario. 
The credible interval fails to capture the truth in two situations, when the true probability is very close to zero, or it is very close to one. 
The former case corresponds to classes with very low success probability, resulting in very few observations to estimate. 
The latter is underestimated as a consequence of the former since the probabilities add up to one.

The results produced by our method are mostly stable and consistent across all synthetic data sets. 
There are, however, a few cases of incorrect cluster assignments, resulting in some outliers in each boxplot. 
Note that if an incorrect cluster assignment takes place, the probabilities of all input-response combinations are affected by that. 
For example, if a component of $ M_{1}$ is wrongly assigned to $M_{2}$, then not only the probabilities of input-output combinations in $M_{1}$ and $M_{2}$ are affected, since the probabilities add up to one, those of $S_{1}$ and $S_{2}$ are also affected.    

In estimating the probabilities, the overall mean squared error, 
i.e., the mean squared difference of the estimated and the true probabilities taking all combinations of $(d,s,i,t)$ into account, came out to be $0.0028$.
Figure \ref{fig: simulation_boxplots} provides a detailed description of the estimation of the probabilities for two input categories (one from each similarity group). 
As described for the individual simulation results, there are cases of under-estimation of the probabilities which are close to one, 
and consequently, over-estimation of the probabilities close to zero. 
However, the amount of departure from the true probability in each case is very small which can also be seen in the small overall MSE.

\afterpage{
	\begin{figure}[ht!]
		\centering
		\hskip 0.5cm\includegraphics[width=0.73\textwidth, trim=2cm 0.25cm 1cm 0.25cm]{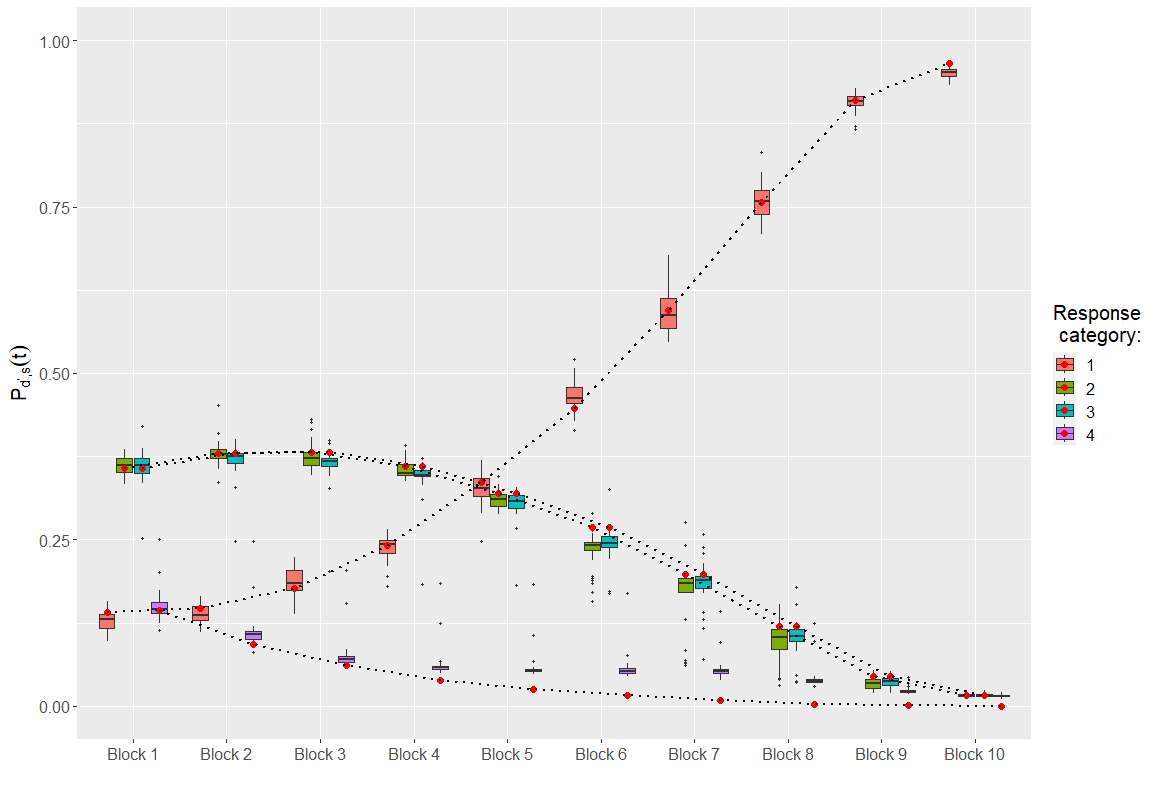}\\
		\hskip 0.5cm\includegraphics[width=0.73\textwidth, trim=2cm 0.25cm 1cm 0.25cm]{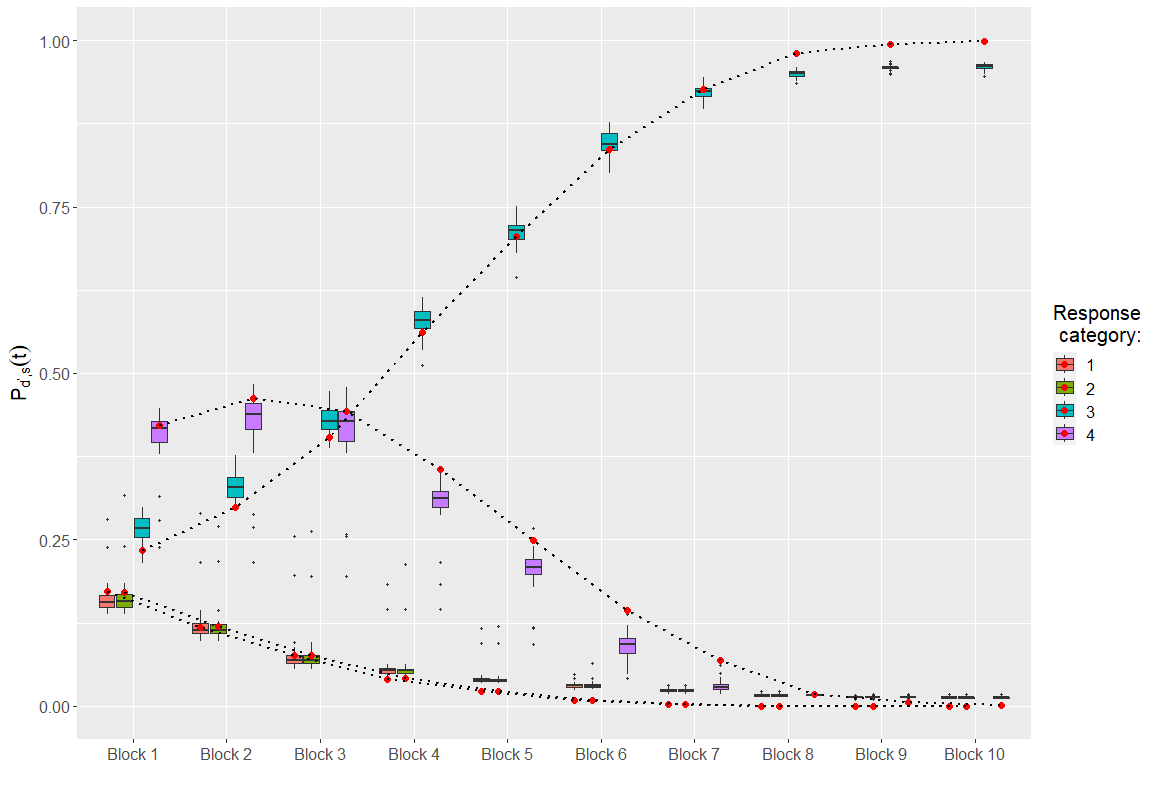}
		\caption{\baselineskip=10pt\small 
			Results of the synthetic data: Boxplots of the estimated probabilities over $50$ simulations, and true probabilities (in red dot) of each block and for two panels, one from each similarity group (panel $T_{1}$ in the top and $T_{3}$ in the bottom).} 
		\label{fig: simulation_boxplots}
	\end{figure}
}

Further, the overall efficiency in identifying the true clustering structure is validated using Rand \citep{Rand1971Objective} and adjusted Rand \citep{Hubert1985Comparing} indices. 
The definitions of Rand and adjusted Rand indices are provided in Section S.6 in the supplementary material. 
The average Rand and adjusted Rand indices for our proposed method over $50$ simulations $0.9105$ and $0.8277$, respectively, indicating high overall efficacy in correctly clustering the probability curves.

\vspace*{-0.5cm}
\section{Applications} \label{sec: application}
\paragraph{Analysis of the PTC1 data set.} 
We present here the analysis of the PTC1 data set described in Section \ref{sec: ptc1} using our proposed longitudinal inverse-probit mixed model. 
We first demonstrate the performance of the proposed method in estimating the probabilities associated with different $(d,s)$ pairs. 
Figure \ref{fig: PTC1_estimated_prob} shows the $95\%$ credible intervals for the estimated probabilities for different input tones, along with the average proportions of times an input tone was classified into different tone categories across subjects.
The latter serves as the empirical estimate of the probabilities.

\afterpage{
	\begin{figure}[ht!]
		\centering
		\hskip 0.5cm\includegraphics[width=0.75\textwidth, trim=2cm 0.25cm 1cm 0.25cm]{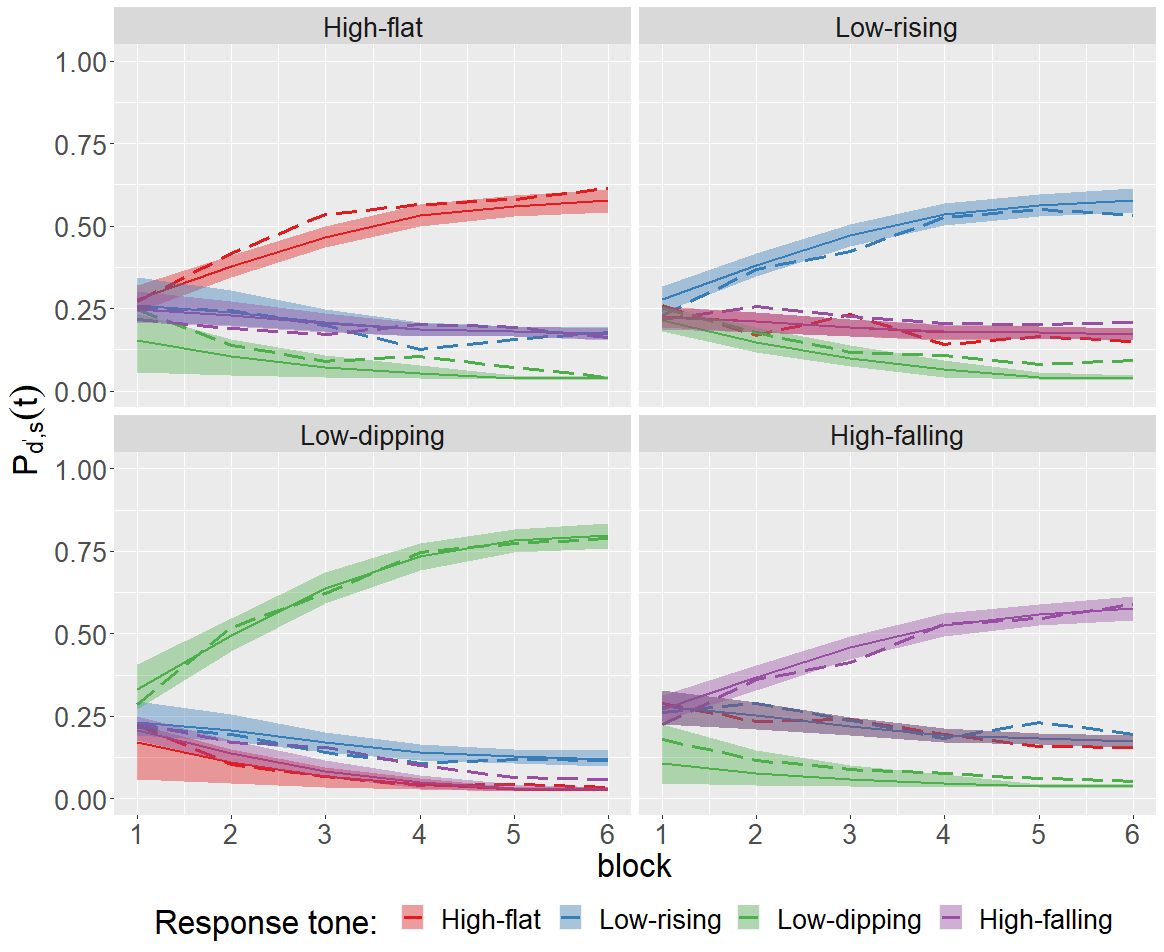}
		\caption{\baselineskip=10pt\small
			Results for PTC1 data: Estimated probability trajectories compared with average proportions of times an input tone was classified into different tone categories across subjects (in dashed line). The means across subjects are indicated by thick lines and the shaded regions indicate corresponding $95\%$ coverage regions.
		}
		\label{fig: PTC1_estimated_prob}
	\end{figure}
}

We observe that except for the input-response combination (1,1) in block 3 and some cases with a low number of data points, 
the $95\%$ credible intervals include the corresponding empirical probabilities. 
An explanation of the occasional under-performance is given later in this section.

Next, we examine the clusters identified by the proposed model.
Apart from the two clusters obtained for the success combinations $(d=s)$, 
three clusters are additionally identified in the incorrect input-response combinations $(d\neq s)$. 
The clusters of success combinations are 
{\small $S_{1}=\{(1,1),(2,2),(4,4)\}$} and {\small $S_{2}=\{(3,3)\}$}, 
and of wrong allocations are 
{\small$M_{1}=\{(1,2),(1,4),(2,1),(2,4),(3,2),(4,1),(4,2)\}$}, {\small $M_{2}=\{(1,3),(2,3),(3,4),(4,3) \}$}, and {\small $M_{3}=\{(3,1)\}$}. 
Figure \ref{fig: PTC1_network} shows the input-response tone combinations color-coded as per cluster identity, 
and the proportion of times each pair of input-response tone combinations appeared in the same cluster after burnin. 
Figure \ref{fig: PTC1_network} indicates that, while the clusters $S_{1}, S_{2}, M_{1}$ are stable, 
there is some instability among the other two clusters, namely $M_{2}$ and $M_{3}$.

\afterpage{
	\begin{figure}[ht!]
		\centering
		\includegraphics[scale=.6,trim=0cm 2cm 3cm 2.25cm,clip=TRUE]{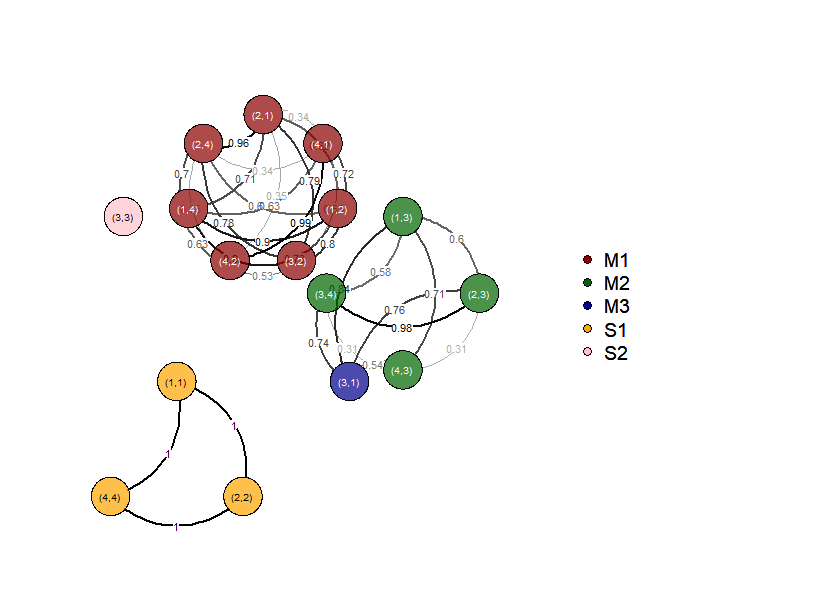}
		\caption{\baselineskip=10pt\small
			Results for PTC1 data: Network plot of similarity groups showing the intra and inter-cluster similarities of tone recognition problems. 
			Each node is associated with a pair indicating the input-response tone category, $(s,d)$. 
			The number associated with each edge indicates the proportion of times the pair in the two connecting nodes appeared in the same cluster after burnin.
		}
		\label{fig: PTC1_network}
	\end{figure}
}

{\it Key findings.} The clustering structure reveals that the low-dipping ($T_{3}$) response trajectories are different from the other three response categories. While for correct input-output tone combinations, $S_{2}$ forms a separate singleton cluster, for incorrect combinations, $M_{2}$ contains all the low-dipping trajectories, indicating their similarities across the panels. 
Also for $T_{3}$, faster increase of the probabilities of correct identification, as well as faster decay of probabilities of incorrect identification indicate that $T_{3}$ is easily distinguishable from other alternatives.

On the other hand, the trajectories of high-flat ($T_{1}$), low-rising ($T_{2}$) and high-falling ($T_{4}$) response categories are quite similar across panels. While for correct input-response combinations, these three form the cluster $S_{1}$, the corresponding incorrect tone combinations are clustered in $M_{1}$. 
The slower rise of the observed empirical probabilities for the elements in $S_{1}$ and the slower decay of the same for $M_{1}$ indicate that $T_{1}$, $T_{2}$ and $T_{4}$ are difficult to distinguish.
However, in block 3 the empirical probabilities of correct input-response combinations differ moderately. 
While $T_{2}$ and $T_{4}$ show a relative drop in the empirical probabilities at block 3, $T_{1}$ shows a sudden pick in the same. 
%Overall similarities lead the model to all these three correct input-response combinations to be assigned to $S_{1}$. However, 
This local dissimilarity of the trajectories at block 3, leads to a departure of the empirical probability of $T_{1}$ from the estimated credible band.

Next, we consider the results concerning the estimation of the drift parameters $\mu_{d^{\prime},s}^{(i)}(t)$. 
As discussed in Section \ref{sec: prop approach}, 
given the identifiability constraints, 
the estimates of $\mu_{d^{\prime},s}^{(i)}(t)$ can only be interpreted on a relative scale. 
Figure \ref{fig: PTC1_population_effect} shows the posterior mean trajectories and associated $95\%$ credible intervals for the projected drift rates. 
%Their relative similarities and differences are consistent with domain knowledge. 

\afterpage{
	\begin{figure}[ht!]
		\centering
		\hskip 0.5cm\includegraphics[width=0.75\textwidth, trim=2cm 0.25cm 1cm 0.25cm]{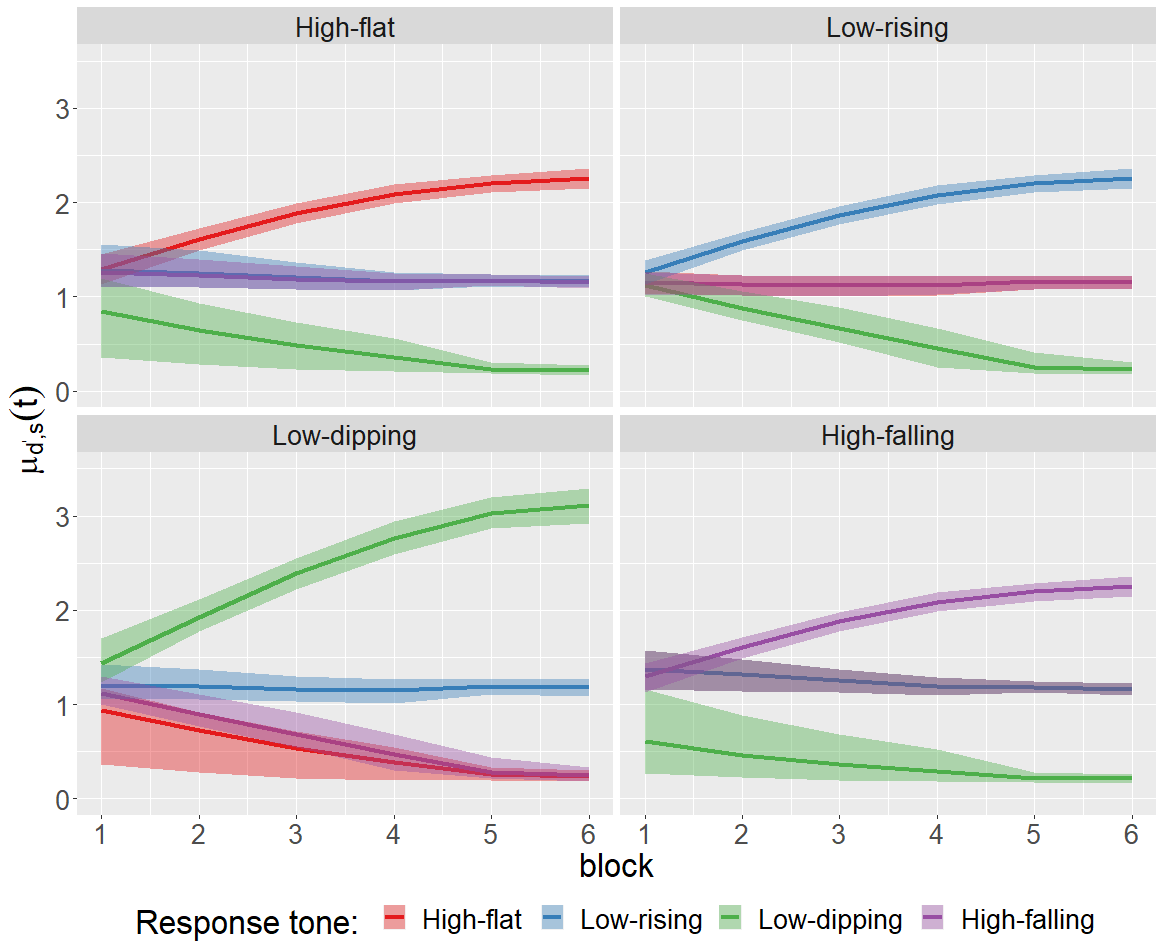}
		\caption{\baselineskip=10pt\small Results for PTC1 data: Estimated posterior mean trajectories of the population level drifts $\mu_{d^{\prime},s}(t)$ for the proposed model. 
			The shaded areas represent the corresponding $95\%$ point-wise credible intervals.
		}
		\label{fig: PTC1_population_effect}
	\end{figure}
}

Importantly, our proposed mixed model also allows us to assess individual-specific parameter trajectories.  
Figure \ref{fig: PTC1_random_effect} shows the posterior mean trajectories and the associated $95\%$ credible intervals 
for the drift rates $\mu_{d^{\prime},s}^{(i)}$ estimated by our method for the different success combinations $(d^{\prime},s)$ for two participants 
- one with the best accuracy averaged across all blocks, and the other with the worst accuracy averaged across all blocks.
These results suggest significant individual-specific heterogeneity. 
For the well-performing participant, the drift parameters are much higher than those for the poorly performing individual, indicating their ability to more quickly accumulate evidence compared to the poorly-performing adult.
These differences persisted over all blocks with a small gradual increase over time.

\afterpage{
	\begin{figure}[ht!]
		\centering
		\hskip 0.5cm\includegraphics[width=0.75\textwidth, trim=2cm 0.25cm 1cm 0.25cm]{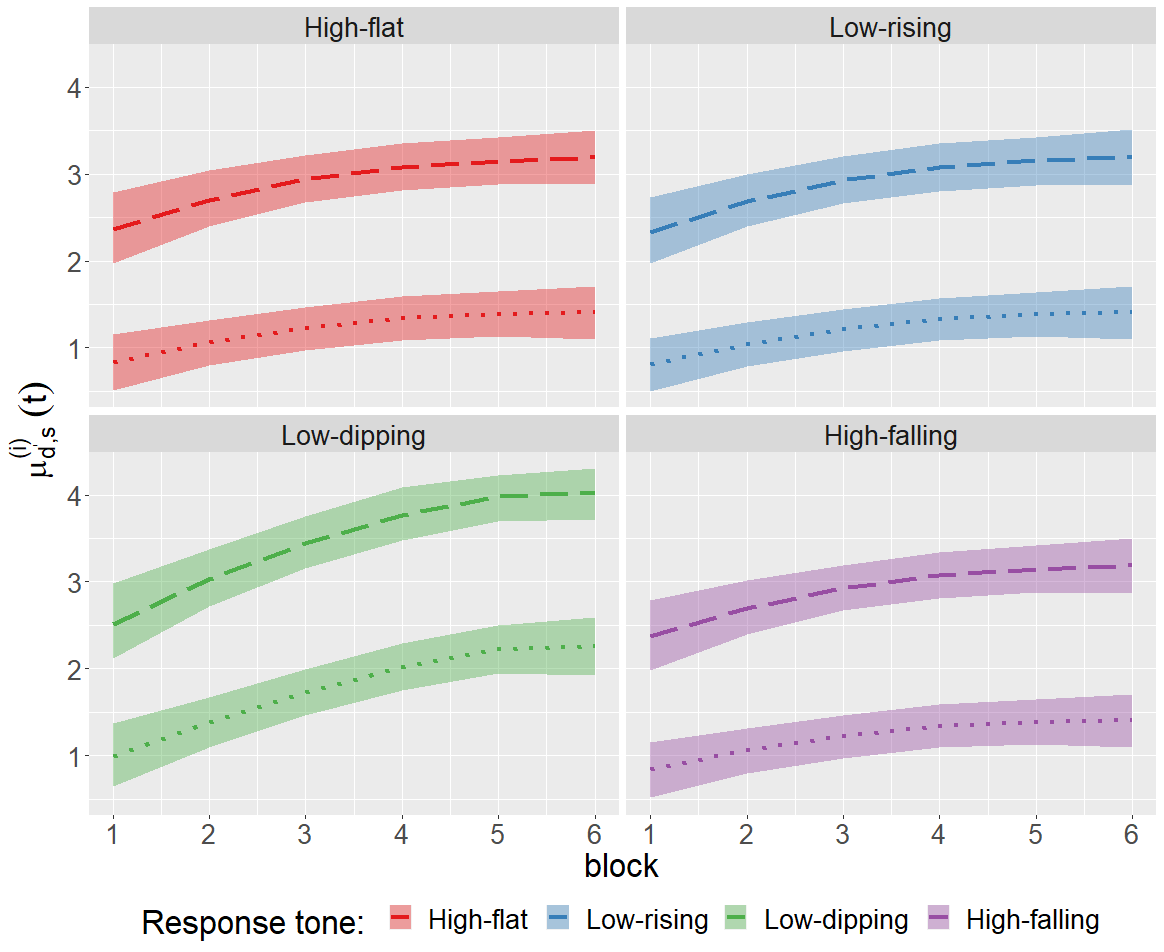}
		\caption{\baselineskip=10pt\small Results for PTC1 data: Estimated posterior mean trajectories for individual specific drifts $\mu_{d^{\prime},s}^{(i)}(t) = \exp \{ f_{d^{\prime},s}(t) + u_{C}^{(i)}(t) \}$ 
			for successful identification $(d^{\prime}=s)$ for two different participants 
			- one performing well (dashed line) and one performing poorly (dotted line).
			The shaded areas represent the corresponding $95\%$ point-wise credible intervals.
		}
		\label{fig: PTC1_random_effect}
	\end{figure}
}

\paragraph{Analysis of benchmark data.} 
To validate the proposed method, 
we also analyzed tone learning data which, in addition to response accuracies, 
included accurate measurements of the response times. 
It was previously analyzed in \cite{paulon2021bayes} using the drift-diffusion model (\ref{eq: inv-Gaussian likelihood}) 
which allowed inference on both the drift and the boundary parameters. 
For our analysis with the method proposed here, however, 
we ignored the response times. 
We observed that the estimates of the drifts produced by our proposed methodology match well with the estimates obtained by \cite{paulon2021bayes}. 
A description of this `benchmark' data set and other details of our analyses are provided in Section S.5 of the supplementary material.

\section{Discussion, Conclusion, Broader Utility, and Future Work} \label{sec: discussion}

\paragraph{Summary.} 
In this article, we developed a novel longitudinal inverse-probit mixed categorical probability model.  
Our research was motivated by category-learning experiments where scientists are interested in using drift-diffusion models to understand 
how the decision-making mechanisms evolve 
as the participants get more training and experience. 
However, unlike traditional drift-diffusion analyses which require data on both response categories and response times,  
we only had usable records of response categories but no response times. 
To our knowledge, biologically interpretable latent drift-diffusion process-based categorical probability models  
had never been considered for such scenarios in the literature before. 
We addressed this need. 
Building on a previous work on longitudinal drift-diffusion mixed joint models for response categories and response times 
but now integrating out the response times, 
we obtained a new class of category probability models which we referred to here as the inverse-probit model. 
We explored parameter recoverability in such models, 
showing, in particular, that the offset parameters can not be recovered 
and drifts and boundaries both can not be recovered from data only on response categories. 
In our analyses, we thus focused on estimating the biologically more important drift parameters but kept the offsets and the boundaries fixed. 
We showed that with careful domain knowledge informed choices for the boundaries, 
the general trajectories of the drift parameters can be recovered by our proposed approach even in the complete absence of response times.

%#################################################################

\paragraph{Conclusion.} 
Overall, when it comes to making scientific inferences about drift-diffusion model parameters 
in the absence of data on response times, 
our work implies a mixed promise. 
On the downside, 
our work shows that 
the detailed interplay between drifts and boundaries cannot be captured. 
On the positive side, our results also suggest that, 
with our carefully designed model, 
and the fixed value of the boundary parameters appropriately chosen by experts, 
the general longitudinal trends in the drifts can still be estimated well. %from data only on response categories. 
Caution should still be exercised not to over-interpret the results.

\paragraph{Broader utility in auditory neuroscience.} 
The proposed model, we believe, has significant implications for auditory neuroscience. 
We focused here specifically on a pupillometry study for which the experimental paradigms need to be adapted to prioritize slow pupillary response, 
rendering the behavioral response times useless. 
However, as discussed in the Introduction, 
there could be many other situations where usable data on response times may not be available. 
The proposed model can be useful in such scenarios to understand the perceptual mechanisms underlying auditory decision-making. 
%The implementation of the proposed mixed model is the first instance to demonstrate promising results for understanding the neural mechanisms underlying auditory decision-making under such scenarios. 
%Long-term implications of this model would allow researchers to study clinical populations with hearing impairments to better understand how rehabilitative approaches (i.e., hearing aids and cochlear implants) may improve decision-making criteria and listening effort, as well as how adult listeners learn a second language.

\paragraph{Broader utility beyond auditory neuroscience.}
While we focused here on studying auditory category learning, the method proposed is applicable to other domains of behavioral neuroscience research studying categorical decision-making when the response times measurements are either not available or not reliable.

\paragraph{Broader utility in statistics.}
On the statistical side, the projection-based approach proposed here to impose non-standard identifiability conditions and address clustering problems within and between different panels is not restricted to inverse-probit models introduced here. 
They can be easily adapted to other classes of generalized linear models such as the widely popular logit and probit models and hence may also be
of interest to a much broader statistical audience.

\paragraph{Future directions:} 
The models and the analyses of the PTC1 data set presented here excluded the pupillometry measurements themselves. 
An important and challenging problem being pursued separately elsewhere is to see how those measurements relate to 
drift-diffusion model parameters.

%\iffalse
\newcommand{\Appendix}{\appendix\def\thesection{\normalsize{Appendix~\Alph{section}}}\def\thesubsection{\Alph{section}.\arabic{subsection}}}
\section*{Appendix}
\begin{appendix}
\Appendix
\renewcommand{\theequation}{A.\arabic{equation}}
\setcounter{equation}{0}
\baselineskip=14pt

\section{\normalsize Proof of Lemma \ref{lem1}} \label{app: lem1}  \label{app: lem1}
\begin{proof}
	It is easy to check that the offset parameters $\delta_{s}$ are not identifiable since 
	\vspace{-4ex}\\
	\bse
	&& P(d \mid s,\delta_{s},\bmu_{1:d_{0},s}, \bb_{1:d_{0},s}) = \int_{\delta_{s}}^{\infty} g(\tau \mid \delta_{s},\mu_{d,s},b_{d,s}) \prod_{d^{\prime} \neq d} \left\{1 - G(\tau \mid \delta_{s},\mu_{d^{\prime},s},b_{d^{\prime},s})\right\} d\tau \\
	&& = \int_{0}^{\infty} g(\tau \mid 0,\mu_{d,s},b_{d,s}) \prod_{d^{\prime} \neq d} \left\{1 - G(\tau \mid 0,\mu_{d^{\prime},s},b_{d^{\prime},s})\right\} d\tau 
	= P(d \mid s, 0,\bmu_{1:d_{0},s},\bb_{1:d_{0},s}).
	\ese
	\vspace{-4ex}

	Next we will show that the drift parameters and decision boundaries are not separately identifiable, even if we fix offset parameters to a constant.
	
	First note that equation (\ref{eq: multinomial inv-probit}) can also be represented as 
	\vspace{-4ex}\\
	\be
	\int_{\delta_{s}}^{\infty} \ldots \int_{\delta_{s}}^{\infty}  \prod_{d^{\prime}\neq d} g(\tau_{d^{\prime}}  \mid \btheta_{d^{\prime},s}) \int_{\delta_{s}}^{\displaystyle\wedge_{d\neq d^{\prime}}\tau_{d^{\prime}} } g(\tau_{d}  \mid \btheta_{d,s}) d\tau_{d} \prod_{d^{\prime}\neq d} d\tau_{d^{\prime}} . \label{eqn:Lm1_1}
	\ee
	\vspace{-4ex}
	
	First observe that $\tau^{\star}=\wedge_{d^{\prime}\neq d}\tau_{d^{\prime}}  =\tau_{-1}^{\star}\wedge \tau_{1}$, where $\tau_{-1}^{\star}=\wedge_{d^{\prime}\neq \{1,d\}}\tau_{d^{\prime}}$.
	Thus the integral above can be written as
	\vspace{-4ex}\\
	\bse
	&&\int_{\delta_{s}}^{\infty} \ldots \int_{\delta_{s}}^{\infty} \prod_{d^{\prime}\neq \{1,d\}} g(\tau_{d^{\prime}} \mid \btheta_{d^{\prime},s}) \left\{ \int_{\delta_{s}}^{\infty} g(\tau_{1} \mid \btheta_{1,s}) \int_{\delta_{s}}^{\tau_{-1}^{\star} \wedge \tau_{1} } g(\tau_{d}  \mid \btheta_{d,s}) d\tau_{d}\right\} \prod_{d^{\prime}\neq \{1,d\}} d\tau_{d^{\prime}}\\
	&& =\int_{\delta_{s}}^{\infty} \ldots \int_{\delta_{s}}^{\infty} \prod_{d^{\prime}\neq \{1,d\}} g(\tau_{d^{\prime}} \mid \btheta_{d^{\prime},s}) \left\{ \int_{\delta_{s}}^{\tau_{-1}^{\star}} g(\tau_{d}  \mid \btheta_{d,s}) \int_{\tau_{d}}^{\infty } g(\tau_{1} \mid \btheta_{1,s})  d\tau_{1} d\tau_{d} \right\} \prod_{d^{\prime}\neq \{1,d\}} d\tau_{d^{\prime}}. 
	\ese
	\vspace{-2ex}\\
	Proceeding sequentially one can show that the integral above is the same as in (\ref{eq: multinomial inv-probit}).
	
	Using the above we express the probability in (\ref{eq: multinomial inv-probit}) as in (\ref{eqn:Lm1_1}). As the offset parameter $\delta_{s}$ is already shown to be not identifiable, we need to fix the same. Without loss of generality, we fix the offset parameter at $0$.
	The probability density function of inverse Gaussian distribution, with parameters $\btheta_{d^{\prime},s}=(\mu_{d^{\prime},s},b_{d^{\prime},s})$ evaluated at $\tau_{d^{\prime}}$, $g(\tau_{d^{\prime}}\mid \btheta_{d^{\prime},s})$ can be obtained from (\ref{eq: drift-diffusion 1}) by replacing $\delta_{s}=0$ and $d=d^{\prime}$.
	
	Consider the transformation of $\tau_{d^{\prime}}$ to $\tau_{d^{\prime}}^{\star}$ as $\tau_{d^{\prime}}=c^2\tau_{d^{\prime}}^{\star}$,
	for some constant $c>0$, and for all $d^{\prime}$. 
	Further, define $b_{d^{\prime},s}^{\star}=b_{d^{\prime},s}/c$ and $\mu_{d^{\prime},s}^{\star}=c\mu_{d^{\prime},s}$, for all $d^{\prime}$.
	Then observe that
	\vspace{-4ex}\\
	\bse
	g(\tau_{d^{\prime}}\mid \btheta_{d^{\prime},s}) d\tau_{d^{\prime}}= (2\pi)^{-1/2} b_{d^{\prime},s}^{\star} (\tau_{d^{\prime}}^{\star})^{-3/2}\exp \left\{-(2\tau_{d^{\prime}}^{\star})^{-1} \left( b_{d^{\prime},s}^{\star} -\mu_{d^{\prime},s}^{\star} \tau_{j}^{\star} \right)^{2} \right\}d\tau_{d^{\prime}}^{\star}=g(\tau_{d^{\prime}}^{\star}\mid \btheta_{d^{\prime},s}^{\star} ) d\tau_{d^{\prime}}^{\star},
	\ese
	\vspace{-4ex}\\
	\noindent where $g(\tau_{d^{\prime}}^{\star}\mid \btheta_{d^{\prime},s}^{\star} )$  is the pdf of inverse Gaussian distribution with parameters $\mu_{d^{\prime},s}^{\star}$ and $b_{d^{\prime},s}^{\star}$,
	evaluated at the point $\tau_{d^{\prime}}^{\star}$.
	
	Applying the transformation on $\tau_{d^{\prime}}$ for all $d^{\prime}$ we get that the integral in (\ref{eqn:Lm1_1}) with $\delta_{s}=0$ is same as 
	\vspace{-4ex}\\
	\bse
	\int_{0}^{\infty} \ldots \int_{0}^{\infty}  \prod_{d^{\prime}\neq d} g(\tau_{d^{\prime}}^{\star}  \mid \btheta_{d^{\prime},s}^{\star}) \int_{0}^{\displaystyle\wedge_{d^{\prime}\neq d}\tau_{d^{\prime}}^{\star} } g(\tau_{d}^{\star}  \mid \btheta_{d,s}^{\star}) d\tau_{d}^{\star} \prod_{d^{\prime}\neq d} d\tau_{d^{\prime}}^{\star}.
	\ese
	\vspace{-2ex}\\
	\noindent As $c$ is arbitrary, this shows that the drifts and boundaries are not separately estimable.
\end{proof}

\section{\normalsize Proof of Theorem \ref{thm:injection}}   \label{app: thm1}
\begin{proof}
	Let $\bP (\bmu_{1:d_{0},s})=\{p_{1}(\bmu_{1:d_{0},s}), \dots, p_{d_{0}}(\bmu_{1:d_{0},s}) \}\trans$ be the function, 
	given by (\ref{eq: mod1b}), 
	from $\S_{0,k}$ to unit probability simplex $\Delta^{d_{0}-1}$.
	For notational simplicity, we write $\bmu_{1:d_{0},s} = \bmu = (\mu_{1}, \dots, \mu_{d_{0}})\trans$. We first find the matrix of partial derivative $\nabla \bP$ with respect to $\bmu$. 
	
	For $\bmu\in \S_{0,k}$, $ {\bf 1}^{T} \bmu=k$, and hence the probability reduces to
	\vspace{-2ex}\\
	\be
	p_{d}\left(\bmu\right) = \displaystyle\frac{\left(be^{b}\right)^{d_{0}}}{  (2 \pi)^{d_{0}/2}}\int_{0}^{\infty} \int_{\tau_{d}}^{\infty} \cdots \int_{\tau_{d}}^{\infty} 
	|\btau|^{-3/2} \exp \left\{-\frac{1}{2} \left( {\bf 1}^{T} \btau^{-1} {\bf 1}  + \bmu^{T} \btau {\bmu} \right) \right\}
	d\btau_{-d} d\tau_{d}, \notag
	\ee
	\vspace{-2ex}\\
	\noindent for $d=1,\ldots, d_{0}$, where $\btau=\diag(\tau_{1} , \ldots, \tau_{d_{0}})$, and $\btau_{-d}$ is the sub-vector of $\btau$ excluding the $d$-th element. 
	Next, differentiating $p_{d}\left(\bmu\right)$ with respect to $\bmu$, we get
	\vspace{-4ex}\\
	\be
	\frac{\partial{p_{d}\left(\bmu\right)}}{\partial \bmu} &= & \displaystyle\frac{\left(be^{b}\right)^{d_{0}}}{  (2 \pi)^{d_{0}/2}}\int_{0}^{\infty} \int_{\tau_{d}}^{\infty} \cdots \int_{\tau_{d}}^{\infty} 
	|\btau|^{-3/2} \left(-\btau \bmu \right)\exp \left\{-\frac{1}{2} \left( {\bf 1}^{T} \btau^{-1} {\bf 1}  + \bmu^{T} \btau {\bmu} \right) \right\}
	d\btau_{-d} d\tau_{d},  \notag\\
	&=& \begin{bmatrix} \mu_{1} \eta_{2}~ & \cdots & ~\mu_{d-1} \eta_{2} & \mu_{d} \eta_{1}  & \mu_{d+1} \eta_{2}~ & \cdots & ~ \mu_{d_{0}} \eta_{2} \end{bmatrix}^{T}, \hspace{1.925 in}\notag
	\ee
	\vspace{-4ex}\\
	\noindent where $ \eta_{1} = - E\left\{ \tau_{1} \mathbb{I}\left( \tau_{2} > \tau_{1}, \cdots , \tau_{d_{0}}>\tau_{1}\right)  \left| \bmu \right. \right\}$, and $ \eta_{2} = -E\left\{ \tau_{2} \mathbb{I}\left( \tau_{2} > \tau_{1}, \cdots , \tau_{d_{0}}>\tau_{1}\right)  \left| \bmu \right. \right\}$, and $\mathbb{I}(A)$ is the indicator function of the event $A$. Here the expectation is considered under the joint distribution of $\left(\tau_{1}, \ldots, \tau_{d}\right)$, which is independent inverse Gaussian. Clearly $\eta_{1}>\eta_{2}>0$.
	
	From the above derivation it is easy to obtain that
	\vspace{-4ex}\\
	\be
	\nabla \bP \left( \bmu \right) = \begin{bmatrix} \mu_{1} \eta_{1} & \mu_{2} \eta_{2} & \cdots & \mu_{d_{0}} \eta_{2} \\
		\mu_{1} \eta_{2} & \mu_{2} \eta_{1} & \cdots & \mu_{d_{0}} \eta_{2} \\
		\vdots &  \vdots & \cdots & \vdots \\
		\mu_{1} \eta_{2} & \mu_{2} \eta_{2} & \cdots & \mu_{d_{0}} \eta_{1}
	\end{bmatrix} = \bM \left\{\left(\eta_{1}-\eta_{2}\right)I+\eta_{2} {\bf 1} {\bf 1}^{T} \right\}, \notag 
	\ee
	\vspace{-2ex}\\
	\noindent where $\bM=\diag \left( \mu_{1}, \ldots, \mu_{d_{0}}\right)$.
	
	Now, suppose there exists $\bmu$ and $\bnu$ in $\S_{k}$ such that $\bmu\neq \bnu$ and $\bP\left(\bmu\right) = \bP\left(\bnu\right)$. Define $\bgamma: [0,1] \rightarrow \mathbb{R}^{d_{0}}$ such that $\bgamma (t)= \bmu + t \left( \bnu - \bmu \right)$, $t\in [0,1]$. Further, define $h(t)= \left< \bP \left(\bgamma(t) \right) - \bP\left( \bmu\right), \bnu - \bmu\right>$, as the cross product of $\bP \left(\bgamma(t) \right) - \bP\left( \bmu\right)$ and $\bnu-\bmu $. Then $h(1)=h(0)=0$ under the proposition that $\bP\left(\bmu\right) = \bP\left(\bnu\right)$. Therefore, by the Mean Value Theorem, as $\bmu\neq \bnu$, there exists some point $c\in (0,1)$ such that $\left. \partial h(t) / \partial t \right|_{t=c} =0$. 
	Now, 
	\vspace{-4ex}\\
	\be
	\frac{\partial h(t) }{ \partial t} &=&  \sum_{d^{\prime}=1}^{d_{0}} \left(\nu_{d^{\prime}} - \mu_{d^{\prime}} \right) \frac{\partial }{\partial t}\left[ p_{d^{\prime}} \left\{\bgamma (t) \right\} - p_{d^{\prime}} \left( \bmu \right)\right] \notag \\
	&=& \sum_{d^{\prime}=1}^{d_{0}} \left(\nu_{d^{\prime}} - \mu_{d^{\prime}} \right) \left\{ \frac{\partial }{\partial \bgamma } p_{d^{\prime}} \left(\bgamma  \right) \right\}^{T} \frac{\partial \bgamma(t)}{\partial t}   \notag\\
	&=& \left( \bnu -\bmu \right)^{T} \nabla \bP \{\bgamma(t)\} \left( \bnu -\bmu \right) \notag\\
	&=&  \left(\eta_{1}-\eta_{2}\right) \left( \bnu -\bmu \right)^{T} \bGamma(t) \left( \bnu -\bmu \right)  + \eta_{2} \left( \bnu -\bmu \right)^{T} M {\bf 1} {\bf 1}^{T} \left( \bnu -\bmu \right) \notag \\
	&=& \left(\eta_{1}-\eta_{2}\right) \left( \bnu -\bmu \right)^{T} \bGamma(t) \left( \bnu -\bmu \right), \notag
	\ee
	\vspace{-4ex}\\
	as ${\bf 1}^{T} \left( \bnu -\bmu \right)=0$, where $\bGamma(t)=\diag\{ \bgamma(t) \}$.
	
	As every component of $\bmu$ and $\bnu$ is positive, for any $c\in (0,1)$, the matrix $\bGamma(c)$ is positive definite. Further, as $\eta_{1}>\eta_{2}$, $\left. \partial h(t) / \partial t \right|_{t=c} =0$ only if $ \bmu=\bnu$, which contradicts the proposition. 
\end{proof}

\section{\normalsize Algorithm for Minimal Distance Mapping} \label{app: algo mindist}
The problem of finding projection of a point $\bmu$ onto the space $\S_{k,\varepsilon}$ is equivalent to the following non-linear optimization problem:
\vspace{-4ex}\\
\bse
\mathrm{minimize}_{\bw} \|{\bw} -\bmu \|^2 \qquad \mbox{such that} \quad \sum_{i=1}^{d_{0}} w_{i}=k,~ w_{i} \geq \varepsilon. 
\ese
\vspace{-4ex}\\
\citet[Algorithm 1]{Duchi2008Efficient} provides a solution to the problem of projection of a given point $\bmu$ onto the space $\S_{k,\varepsilon}$ for $\varepsilon=0$, which is modified for any given $\varepsilon$ below. 

\begin{algorithm}[ht!] 
	\caption{}
	\label{algo: MCMC}
	\begin{algorithmic}[1]
		\vspace{0.2cm}
		\Algphase{INPUT: A vector $\bmu$, and a pair $(k,\varepsilon)$.}
		\State Sort $\bmu$ into $\bmu^{\star}$ such that the elements of $\bmu^{\star}$ are in descending order.
		\State Find $\rho =\max\left\{j: \mu^{\star}_{j}- j^{-1} \left( \sum_{l=1}^{j}\mu_{l}^{\star} -k\right)>\varepsilon \right\}$.
		\State Define $\theta=\rho^{-1}\left\{ \sum_{l=1}^{j}\mu_{l}^{\star} -k + (k -\rho) \varepsilon\right\}$.
		\Algphase{OUTPUT: ${\bw}$ such that $w_{i}=\max\left\{\mu_{i}-\theta, \varepsilon \right\}$.} 
	\end{algorithmic} 
\end{algorithm}

\section{\normalsize Proof of Lemma \ref{lem3}} \label{app: lem3}
\begin{proof}
	We consider the unconditional distribution of $\tau_{1:d_{0}}$, given the parameters $\mu_{1:d_{0}}$ as the proposal distribution, $g$. Clearly, the proposal distribution $g$ and the target conditional joint distribution $f$ satisfies $f(\tau_{1:d_{0}}|\mu_{1:d_{0}})/g(\tau_{1:d_{0}}|\mu_{1:d_{0}})\leq M$, where $M^{-1}=P\left( \tau_{d} \leq \tau_{1:d_{0}}\right)$.
	Therefore, for any random sample $U\sim U(0,1)$, $f(\tau_{1:d_{0}}|\mu_{1:d_{0}})\geq M U g(\tau_{1:d_{0}}|\mu_{1:d_{0}})$ if the sample satisfies the condition $\tau_{d} \leq \tau_{1:d_{0}}$, and $f(\tau_{1:d_{0}}|\mu_{1:d_{0}})< M U g(\tau_{1:d_{0}}|\mu_{1:d_{0}})$ otherwise. Hence by Lemma 2.3.1 of \cite{Robert2004Monte}, the algorithm above produces samples from the target distribution.
\end{proof}

\end{appendix}

\section*{Supplementary Materials}
The supplementary materials detail the choice of the prior hyper-parameters, the MCMC algorithm used to sample from the posterior and some performance diagnostics, and the analysis of a real benchmark data set. 
Separate files additionally include R programs implementing the longitudinal inverse-probit mixed model developed in this article and the PTC1 data set analyzed in Section \ref{sec: application}.

\iffalse
\section*{Acknowledgments}
This work was supported by the National Science Foundation grant NSF DMS-1953712 and National Institute on Deafness and Other Communication Disorders grants R01DC013315 and R01DC015504, awarded to Sarkar and Chandrasekaran. 
\fi

%\clearpage
%\vspace{1cm}
\baselineskip=14pt
\bibliographystyle{natbib}
\bibliography{Categorical,Projection,Diffusion,FDA_LDA,HMM,HOHMM,MCMC_Latent_Var_Models,neuro}

\begin{thebibliography}{}

\bibitem[Agresti(2018)Agresti]{agresti2018introduction}
Agresti, A. (2018).
\newblock {\em An introduction to categorical data analysis\/}.
\newblock Wiley.

\bibitem[Albert and Chib(1993)Albert and Chib]{Albert1993Bayesian}
Albert, J.~H. and Chib, S. (1993).
\newblock Bayesian analysis of binary and polychotomous response data.
\newblock {\em Journal of the American statistical Association\/}, {\bf 88},
  669--679.

\bibitem[Ashby {\em et~al.}(2003)Ashby, Noble, Filoteo, Waldron, and
  Ell]{ashby2003category}
Ashby, F.~G., Noble, S., Filoteo, J.~V., Waldron, E.~M., and Ell, S.~W. (2003).
\newblock Category learning deficits in parkinson's disease.
\newblock {\em Neuropsychology\/}, {\bf 17}, 115.

\bibitem[Beck(2017)Beck]{Beck2017First}
Beck, A. (2017).
\newblock {\em First-order methods in optimization\/}.
\newblock SIAM.

\bibitem[Bogacz {\em et~al.}(2010)Bogacz, Wagenmakers, Forstmann, and
  Nieuwenhuis]{bogacz2010neural}
Bogacz, R., Wagenmakers, E.-J., Forstmann, B.~U., and Nieuwenhuis, S. (2010).
\newblock The neural basis of the speed-accuracy tradeoff.
\newblock {\em Trends in Neurosciences\/}, {\bf 33}, 10--16.

\bibitem[Borooah(2002)Borooah]{borooah2002logit}
Borooah, V.~K. (2002).
\newblock {\em Logit and probit: ordered and multinomial models\/}.
\newblock Sage.

\bibitem[Brody and Hanks(2016)Brody and Hanks]{brody2016neural}
Brody, C.~D. and Hanks, T.~D. (2016).
\newblock Neural underpinnings of the evidence accumulator.
\newblock {\em Current Opinion in Neurobiology\/}, {\bf 37}, 149--157.

\bibitem[Brown and Heathcote(2008)Brown and Heathcote]{brown2008simplest}
Brown, S.~D. and Heathcote, A. (2008).
\newblock The simplest complete model of choice response time: Linear ballistic
  accumulation.
\newblock {\em Cognitive Psychology\/}, {\bf 57}, 153--178.

\bibitem[Burgette and Nordheim(2012)Burgette and Nordheim]{burgette2012trace}
Burgette, L.~F. and Nordheim, E.~V. (2012).
\newblock The trace restriction: {A}n alternative identification strategy for
  the {B}ayesian multinomial probit model.
\newblock {\em Journal of Business \& Economic Statistics\/}, {\bf 30},
  404--410.

\bibitem[Burgette {\em et~al.}(2021)Burgette, Puelz, and Hahn]{Burgette2021A}
Burgette, L.~F., Puelz, D., and Hahn, P.~R. (2021).
\newblock A symmetric prior for multinomial probit models.
\newblock {\em Bayesian Analysis\/}, {\bf 16}, 991--1008.

\bibitem[Cavanagh {\em et~al.}(2011)Cavanagh, Wiecki, Cohen, Figueroa, Samanta,
  Sherman, and Frank]{cavanagh2011subthalamic}
Cavanagh, J.~F., Wiecki, T.~V., Cohen, M.~X., Figueroa, C.~M., Samanta, J.,
  Sherman, S.~J., and Frank, M.~J. (2011).
\newblock Subthalamic nucleus stimulation reverses mediofrontal influence over
  decision threshold.
\newblock {\em Nature Neuroscience\/}, {\bf 14}, 1462.

\bibitem[Chandrasekaran {\em et~al.}(2014)Chandrasekaran, Yi, and
  Maddox]{chandrasekaran2014dual}
Chandrasekaran, B., Yi, H.-G., and Maddox, W.~T. (2014).
\newblock Dual-learning systems during speech category learning.
\newblock {\em Psychonomic Bulletin \& Review\/}, {\bf 21}, 488--495.

\bibitem[Chandrasekaran {\em et~al.}(2016)Chandrasekaran, Yi, Smayda, and
  Maddox]{chandrasekaran2016effect}
Chandrasekaran, B., Yi, H.-G., Smayda, K.~E., and Maddox, W.~T. (2016).
\newblock Effect of explicit dimensional instruction on speech category
  learning.
\newblock {\em Attention, Perception, \& Psychophysics\/}, {\bf 78}, 566--582.

\bibitem[Chhikara(1988)Chhikara]{chhikara1988inverse}
Chhikara, R. (1988).
\newblock {\em The inverse Gaussian distribution: Theory, methodology, and
  applications\/}.
\newblock CRC Press.

\bibitem[Chib and Greenberg(1998)Chib and Greenberg]{chib1998analysis}
Chib, S. and Greenberg, E. (1998).
\newblock Analysis of multivariate probit models.
\newblock {\em Biometrika\/}, {\bf 85}, 347--361.

\bibitem[Cox and Miller(1965)Cox and Miller]{cox1965theory}
Cox, D.~R. and Miller, H.~D. (1965).
\newblock {\em The theory of stochastic processes\/}.
\newblock CRC Press.

\bibitem[de~Boor(1978)de~Boor]{de1978practical}
de~Boor, C. (1978).
\newblock {\em A practical guide to splines\/}.
\newblock Springer-Verlag.

\bibitem[Deo(2018)Deo]{Deo2018Algebraic}
Deo, S. (2018).
\newblock {\em Algebraic topology\/}, volume~27 of {\em Texts and Readings in
  Mathematics\/}.
\newblock Hindustan Book Agency, New Delhi.

\bibitem[Ding and Gold(2013)Ding and Gold]{ding2013basal}
Ding, L. and Gold, J.~I. (2013).
\newblock The basal ganglia's contributions to perceptual decision making.
\newblock {\em Neuron\/}, {\bf 79}, 640--649.

\bibitem[Duchi {\em et~al.}(2008)Duchi, Shalev-Shwartz, Singer, and
  Chandra]{Duchi2008Efficient}
Duchi, J., Shalev-Shwartz, S., Singer, Y., and Chandra, T. (2008).
\newblock Efficient projections onto the l 1-ball for learning in high
  dimensions.
\newblock In {\em Proceedings of the 25th International Conference on Machine
  Learning\/}, pages 272--279.

\bibitem[Dufau {\em et~al.}(2012)Dufau, Grainger, and Ziegler]{dufau2012say}
Dufau, S., Grainger, J., and Ziegler, J.~C. (2012).
\newblock How to say ``no'' to a nonword: A leaky competing accumulator model
  of lexical decision.
\newblock {\em Journal of Experimental Psychology: Learning, Memory, and
  Cognition\/}, {\bf 38}, 1117.

\bibitem[Dunson and Neelon(2003)Dunson and Neelon]{Dunson2003Bayesian}
Dunson, D.~B. and Neelon, B. (2003).
\newblock Bayesian inference on order-constrained parameters in generalized
  linear models.
\newblock {\em Biometrics\/}, {\bf 59}, 286--295.

\bibitem[Eilers and Marx(1996)Eilers and Marx]{eilers1996flexible}
Eilers, P.~H. and Marx, B.~D. (1996).
\newblock Flexible smoothing with b-splines and penalties.
\newblock {\em Statistical Science\/}, {\bf 11}, 89--102.

\bibitem[Filoteo {\em et~al.}(2010)Filoteo, Lauritzen, and
  Maddox]{filoteo2010removing}
Filoteo, J.~V., Lauritzen, S., and Maddox, W.~T. (2010).
\newblock Removing the frontal lobes: The effects of engaging executive
  functions on perceptual category learning.
\newblock {\em Psychological science\/}, {\bf 21}, 415--423.

\bibitem[Geweke(1992)Geweke]{Geweke1992Evaluating}
Geweke, J. (1992).
\newblock Evaluating the accuracy of sampling-based approaches to the
  calculation of posterior moments.
\newblock In {\em Bayesian statistics, 4 ({P}eniscola, 1991)\/}, pages
  169--193. Oxford Univ. Press, New York.

\bibitem[Glimcher and Fehr(2013)Glimcher and Fehr]{glimcher2013neuroeconomics}
Glimcher, P.~W. and Fehr, E. (2013).
\newblock {\em Neuroeconomics: Decision making and the brain\/}.
\newblock Academic Press.

\bibitem[Gold and Shadlen(2007)Gold and Shadlen]{gold2007neural}
Gold, J.~I. and Shadlen, M.~N. (2007).
\newblock The neural basis of decision making.
\newblock {\em Annual Review of Neuroscience\/}, {\bf 30}, 535--574.

\bibitem[Gunn and Dunson(2005)Gunn and Dunson]{Gunn2005Transformation}
Gunn, L.~H. and Dunson, D.~B. (2005).
\newblock A transformation approach for incorporating monotone or unimodal
  constraints.
\newblock {\em Biostatistics\/}, {\bf 6}, 434--449.

\bibitem[Heekeren {\em et~al.}(2004)Heekeren, Marrett, Bandettini, and
  Ungerleider]{heekeren2004general}
Heekeren, H.~R., Marrett, S., Bandettini, P.~A., and Ungerleider, L.~G. (2004).
\newblock A general mechanism for perceptual decision-making in the human
  brain.
\newblock {\em Nature\/}, {\bf 431}, 859.

\bibitem[Hubert and Arabie(1985)Hubert and Arabie]{Hubert1985Comparing}
Hubert, L. and Arabie, P. (1985).
\newblock Comparing partitions.
\newblock {\em Journal of Classification\/}, {\bf 2}, 193--218.

\bibitem[Johndrow {\em et~al.}(2013)Johndrow, Dunson, and
  Lum]{johndrow2013diagonal}
Johndrow, J., Dunson, D., and Lum, K. (2013).
\newblock Diagonal orthant multinomial probit models.
\newblock In {\em Artificial Intelligence and Statistics\/}, pages 29--38.

\bibitem[Kim {\em et~al.}(2017)Kim, Potter, Craigmile, Peruggia, and
  Van~Zandt]{kim2017bayesian}
Kim, S., Potter, K., Craigmile, P.~F., Peruggia, M., and Van~Zandt, T. (2017).
\newblock A {B}ayesian race model for recognition memory.
\newblock {\em Journal of the American Statistical Association\/}, {\bf 112},
  77--91.

\bibitem[Lau and Green(2007)Lau and Green]{lau2007bayesian}
Lau, J.~W. and Green, P.~J. (2007).
\newblock Bayesian model-based clustering procedures.
\newblock {\em Journal of Computational and Graphical Statistics\/}, {\bf 16},
  526--558.

\bibitem[Leite and Ratcliff(2010)Leite and Ratcliff]{leite2010modeling}
Leite, F.~P. and Ratcliff, R. (2010).
\newblock Modeling reaction time and accuracy of multiple-alternative
  decisions.
\newblock {\em Attention, Perception, \& Psychophysics\/}, {\bf 72}, 246--273.

\bibitem[Llanos {\em et~al.}(2020)Llanos, McHaney, Schuerman, Yi, Leonard, and
  Chandrasekaran]{llanos2020non}
Llanos, F., McHaney, J.~R., Schuerman, W.~L., Yi, H.~G., Leonard, M.~K., and
  Chandrasekaran, B. (2020).
\newblock Non-invasive peripheral nerve stimulation selectively enhances speech
  category learning in adults.
\newblock {\em NPJ science of learning\/}, (1), 1--11.

\bibitem[Lu(1995)Lu]{lu1995degradation}
Lu, J. (1995).
\newblock {\em Degradation processes and related reliability models\/}.
\newblock Ph.D. thesis, McGill University, Montreal, Canada.

\bibitem[McHaney {\em et~al.}(2021)McHaney, Tessmer, Roark, and
  Chandrasekaran]{mchaney2021working}
McHaney, J.~R., Tessmer, R., Roark, C.~L., and Chandrasekaran, B. (2021).
\newblock Working memory relates to individual differences in speech category
  learning: Insights from computational modeling and pupillometry.
\newblock {\em Brain and Language\/}, {\bf 22}, 1--15.

\bibitem[Milosavljevic {\em et~al.}(2010)Milosavljevic, Malmaud, Huth, Koch,
  and Rangel]{milosavljevic2010drift}
Milosavljevic, M., Malmaud, J., Huth, A., Koch, C., and Rangel, A. (2010).
\newblock The drift diffusion model can account for the accuracy and reaction
  time of value-based choices under high and low time pressure.
\newblock {\em Judgment and Decision Making\/}, {\bf 5}, 437--449.

\bibitem[Morris(2015)Morris]{morris2015functional}
Morris, J.~S. (2015).
\newblock Functional regression.
\newblock {\em Annual Review of Statistics and Its Application\/}, {\bf 2},
  321--359.

\bibitem[Parthasarathy {\em et~al.}(2020)Parthasarathy, Hancock, Bennett,
  DeGruttola, and Polley]{parthasarathy2020bottom}
Parthasarathy, A., Hancock, K.~E., Bennett, K., DeGruttola, V., and Polley,
  D.~B. (2020).
\newblock Bottom-up and top-down neural signatures of disordered multi-talker
  speech perception in adults with normal hearing.
\newblock {\em Elife\/}, {\bf 9}, e51419.

\bibitem[Paulon {\em et~al.}(2021)Paulon, Llanos, Chandrasekaran, and
  Sarkar]{paulon2021bayes}
Paulon, G., Llanos, F., Chandrasekaran, B., and Sarkar, A. (2021).
\newblock Bayesian semiparametric longitudinal drift-diffusion mixed models for
  tone learning in adults.
\newblock {\em Journal of the American Statistical Association,\/}, {\bf 116},
  1114--1127.

\bibitem[Peelle(2018)Peelle]{peelle2018listening}
Peelle, J.~E. (2018).
\newblock Listening effort: {H}ow the cognitive consequences of acoustic
  challenge are reflected in brain and behavior.
\newblock {\em Ear and Hearing\/}, {\bf 39}, 204--214.

\bibitem[Purcell(2013)Purcell]{purcell2013neural}
Purcell, B.~A. (2013).
\newblock {\em Neural mechanisms of perceptual decision making\/}.
\newblock Vanderbilt University.

\bibitem[Ramsay and Silverman(2007)Ramsay and Silverman]{ramsay2007applied}
Ramsay, J.~O. and Silverman, B.~W. (2007).
\newblock {\em Applied functional data analysis: {M}ethods and case studies\/}.
\newblock Springer.

\bibitem[Rand(1971)Rand]{Rand1971Objective}
Rand, W.~M. (1971).
\newblock Objective criteria for the evaluation of clustering methods.
\newblock {\em Journal of the American Statistical Association\/}, {\bf 66},
  846--850.

\bibitem[Ratcliff(1978)Ratcliff]{ratcliff1978theory}
Ratcliff, R. (1978).
\newblock A theory of memory retrieval.
\newblock {\em Psychological Review\/}, {\bf 85}, 59.

\bibitem[Ratcliff and McKoon(2008)Ratcliff and McKoon]{ratcliff2008diffusion}
Ratcliff, R. and McKoon, G. (2008).
\newblock The diffusion decision model: Theory and data for two-choice decision
  tasks.
\newblock {\em Neural Computation\/}, {\bf 20}, 873--922.

\bibitem[Ratcliff and Rouder(1998)Ratcliff and Rouder]{ratcliff1998modeling}
Ratcliff, R. and Rouder, J.~N. (1998).
\newblock Modeling response times for two-choice decisions.
\newblock {\em Psychological Science\/}, {\bf 9}, 347--356.

\bibitem[Ratcliff {\em et~al.}(2016)Ratcliff, Smith, Brown, and
  McKoon]{ratcliff2016diffusion}
Ratcliff, R., Smith, P.~L., Brown, S.~D., and McKoon, G. (2016).
\newblock Diffusion decision model: Current issues and history.
\newblock {\em Trends in Cognitive Sciences\/}, {\bf 20}, 260--281.

\bibitem[Reetzke {\em et~al.}(2018)Reetzke, Xie, Llanos, and
  Chandrasekaran]{reetzke2018tracing}
Reetzke, R., Xie, Z., Llanos, F., and Chandrasekaran, B. (2018).
\newblock Tracing the trajectory of sensory plasticity across different stages
  of speech learning in adulthood.
\newblock {\em Current Biology\/}, {\bf 28}, 1419--1427.

\bibitem[Roark {\em et~al.}(2021)Roark, Smayda, and
  Chandrasekaran]{roark2021auditory}
Roark, C.~L., Smayda, K.~E., and Chandrasekaran, B. (2021).
\newblock Auditory and visual category learning in musicians and nonmusicians.
\newblock {\em Journal of Experimental Psychology: General\/}.

\bibitem[Robert and Casella(2004)Robert and Casella]{Robert2004Monte}
Robert, C.~P. and Casella, G. (2004).
\newblock {\em Monte {C}arlo statistical methods\/}.
\newblock Springer Texts in Statistics. Springer-Verlag, New York, second
  edition.

\bibitem[Robison and Unsworth(2019)Robison and
  Unsworth]{robison2019pupillometry}
Robison, M.~K. and Unsworth, N. (2019).
\newblock Pupillometry tracks fluctuations in working memory performance.
\newblock {\em Attention, Perception, \& Psychophysics\/}, {\bf 81}, 407--419.

\bibitem[Ross {\em et~al.}(1996)Ross, Kelly, Sullivan, Perry, Mercer, Davis,
  Washburn, Sager, Boyce, and Bristow]{ross1996stochastic}
Ross, S.~M., Kelly, J.~J., Sullivan, R.~J., Perry, W.~J., Mercer, D., Davis,
  R.~M., Washburn, T.~D., Sager, E.~V., Boyce, J.~B., and Bristow, V.~L.
  (1996).
\newblock {\em Stochastic processes\/}.
\newblock Wiley New York.

\bibitem[Rudin(1991)Rudin]{Rudin1991Functional}
Rudin, W. (1991).
\newblock {\em Functional analysis\/}.
\newblock International Series in Pure and Applied Mathematics. McGraw-Hill,
  Inc., New York, second edition.

\bibitem[Schall(2001)Schall]{schall2001neural}
Schall, J.~D. (2001).
\newblock Neural basis of deciding, choosing and acting.
\newblock {\em Nature Reviews Neuroscience\/}, {\bf 2}, 33.

\bibitem[Sen {\em et~al.}(2018)Sen, Patra, and Dunson]{sen2018constrained}
Sen, D., Patra, S., and Dunson, D. (2018).
\newblock Constrained inference through posterior projections.
\newblock {\em arXiv preprint arXiv:1812.05741\/}.

\bibitem[Smayda {\em et~al.}(2015)Smayda, Chandrasekaran, and
  Maddox]{smayda2015enhanced}
Smayda, K.~E., Chandrasekaran, B., and Maddox, W.~T. (2015).
\newblock Enhanced cognitive and perceptual processing: {A} computational basis
  for the musician advantage in speech learning.
\newblock {\em Frontiers in Psychology\/}, pages 1--14.

\bibitem[Smith and Ratcliff(2004)Smith and Ratcliff]{smith2004psychology}
Smith, P.~L. and Ratcliff, R. (2004).
\newblock Psychology and neurobiology of simple decisions.
\newblock {\em Trends in Neurosciences\/}, {\bf 27}, 161--168.

\bibitem[Smith and Vickers(1988)Smith and Vickers]{smith1988accumulator}
Smith, P.~L. and Vickers, D. (1988).
\newblock The accumulator model of two-choice discrimination.
\newblock {\em Journal of Mathematical Psychology\/}, {\bf 32}, 135--168.

\bibitem[Usher and McClelland(2001)Usher and McClelland]{usher2001time}
Usher, M. and McClelland, J.~L. (2001).
\newblock The time course of perceptual choice: The leaky, competing
  accumulator model.
\newblock {\em Psychological Review\/}, {\bf 108}, 550.

\bibitem[Wade(2023)Wade]{wade2023cluster}
Wade, S. (2023).
\newblock Bayesian cluster analysis.
\newblock {\em Philosophical Transactions of the Royal Society A\/}, {\bf 381},
  1--20.

\bibitem[Wang {\em et~al.}(2016)Wang, Chiou, and
  M{\"u}ller]{wang2016functional}
Wang, J.-L., Chiou, J.-M., and M{\"u}ller, H.-G. (2016).
\newblock Functional data analysis.
\newblock {\em Annual Review of Statistics and Its Application\/}, {\bf 3},
  257--295.

\bibitem[Wang {\em et~al.}(1999)Wang, Spence, Jongman, and
  Sereno]{wang1999training}
Wang, Y., Spence, M.~M., Jongman, A., and Sereno, J.~A. (1999).
\newblock Training {A}merican listeners to perceive {M}andarin tones.
\newblock {\em The Journal of the Acoustical Society of America\/}, {\bf 106},
  3649--3658.

\bibitem[Wang {\em et~al.}(2003)Wang, Jongman, and Sereno]{wang2003acoustic}
Wang, Y., Jongman, A., and Sereno, J.~A. (2003).
\newblock Acoustic and perceptual evaluation of {M}andarin tone productions
  before and after perceptual training.
\newblock {\em The Journal of the Acoustical Society of America\/}, {\bf 113},
  1033--1043.

\bibitem[Whitmore and Seshadri(1987)Whitmore and
  Seshadri]{whitmore1987heuristic}
Whitmore, G. and Seshadri, V. (1987).
\newblock A heuristic derivation of the inverse gaussian distribution.
\newblock {\em The American Statistician\/}, {\bf 41}, 280--281.

\bibitem[Winn {\em et~al.}(2018)Winn, Wendt, Koelewijn, and
  Kuchinsky]{winn2018best}
Winn, M.~B., Wendt, D., Koelewijn, T., and Kuchinsky, S.~E. (2018).
\newblock Best practices and advice for using pupillometry to measure listening
  effort: An introduction for those who want to get started.
\newblock {\em Trends in Hearing\/}, {\bf 22}, 1--32.

\bibitem[Zekveld {\em et~al.}(2011)Zekveld, Kramer, and
  Festen]{zekveld2011cognitive}
Zekveld, A.~A., Kramer, S.~E., and Festen, J.~M. (2011).
\newblock Cognitive load during speech perception in noise: {T}he influence of
  age, hearing loss, and cognition on the pupil response.
\newblock {\em Ear and Hearing\/}, {\bf 32}, 498--510.

\end{thebibliography}

%%%%%%%%%%%%%%%%%%%%%%%%%
%%% Supplementary Materials %%%
%%%%%%%%%%%%%%%%%%%%%%%%%

\clearpage\pagebreak\newpage
\newgeometry{textheight=9in, textwidth=6.5in,}
\pagestyle{fancy}
\fancyhf{}
\rhead{\bfseries\thepage}
\lhead{\bfseries SUPPLEMENTARY MATERIAL}

\baselineskip 20pt
\begin{center}
{\LARGE{Supplementary Materials for\\} 
\bf Bayesian Semiparametric Longitudinal\\ 
\vskip -16pt Inverse-Probit Mixed Models\\ 
for Category Learning
}
\end{center}

\setcounter{equation}{0}
\setcounter{page}{1}
\setcounter{table}{1}
\setcounter{figure}{0}
\setcounter{section}{0}
\numberwithin{table}{section}
\renewcommand{\theequation}{S.\arabic{equation}}
\renewcommand{\thesubsection}{S.\arabic{section}.\arabic{subsection}}
\renewcommand{\thesection}{S.\arabic{section}}
\renewcommand{\thepage}{S.\arabic{page}}
\renewcommand{\thetable}{S.\arabic{table}}
\renewcommand{\thefigure}{S.\arabic{figure}}
\baselineskip=15pt

\vspace{0cm}

\begin{center}
Minerva Mukhopadhyay$^{1}$(minervam@iitk.ac.in)\\
Jacie McHaney$^{2}$ (j.mchaney@pitt.edu)\\
Bharath Chandrasekaran$^{2}$(b.chandra@pitt.edu)\\
Abhra Sarkar$^{3}$ (abhra.sarkar@utexas.edu)\\

\vskip 7mm 
$^{1}$Department of of Mathematics and Statistics,\\
Indian Institute of Technology,\\ 
Kanpur, UP 208016, India\\
\vskip 8pt 
$^{2}$Department of Communication Science and Disorders,\\ 
University of Pittsburgh,\\
4028 Forbes Tower, Pittsburgh, PA 15260, USA\\
\vskip 8pt 
$^{3}$Department of Statistics and Data Sciences, \\
University of Texas at Austin,\\ 
105 East 24th Street D9800, Austin, TX 78712, USA\\
\end{center}

\vskip 10mm
The supplementary materials detail the choice of the prior hyper-parameters, the MCMC algorithm used to sample from the posterior and some performance diagnostics, and the analysis of a real benchmark data set. 
Separate files additionally include R programs implementing the longitudinal inverse-probit mixed model developed in this article and the PTC1 data set analyzed in Section 6 in the main paper.

\newpage
\section{Modelling the Drift Parameters} \label{sec:FixedRandom Model}

\subsection{Functional Fixed Effects} \label{sec: fixed effects}
We model the fixed effects functions $f_{x}(t)$ 
using flexible mixtures of B-spline bases \citeplatex{de1978practical} 
that allow them to smoothly vary with time $t$ 
while also depending locally on the indexing variable $x$ as 
\vspace{-6ex}\\
\be
\textstyle f_{x}(t) = \sum_{k=1}^{K} \beta_{x,k} B_{k}(t) = \bB(t)\bbeta_{x}. \label{eq: fixed effects function}
\ee 
\vspace{-4ex}\\
Here $\bB(t) = \{B_{1}(t),\dots, B_{K}(t)\}$ are a set of known locally supported basis functions spanning $[1,T]$, 
$\bbeta_{x} = (\beta_{x,1},\dots,\beta_{x,K})\trans$ are associated unknown coefficients to be estimated from the data. 
Allowing the $\bbeta_{x}$'s to flexibly vary with $x$ can generate widely different shapes for different input-response category combinations.

\iffalse
\begin{figure}[!ht]
	\centering	\includegraphics[width=.95\linewidth]{./Figures/Qudratic_BSplines_2.pdf}
	\caption{\baselineskip=10pt 
		Plot of 9 quadratic B-splines on an interval $[1,T]$ defined by $11$ knot points that divide $[1,T]$ into $K=6$ equal subintervals.
	}
	\label{fig: b-splines}
\end{figure}
\fi

Towards clustering the fixed effects curves, 
we introduce a set of latent variables $z_{x}$ for each input-response category combination $x$ 
with a shared state space $\{1, \dots, z_{\max}\}$ and associated coefficient atoms $\bbeta_{z}^{\star} = (\beta_{z,1}^{\star},\dots,\beta_{z,K}^{\star})\trans$, we let 
\vspace{-4ex}\\
\be
\begin{split}
	& \textstyle (\bbeta_{x} \mid z_{x}=z) = \bbeta_{z}^{\star}, ~~\text{implying}~~\{f_{x}(t) \mid z_{x}=z\} = f_{z}^{\star}(t) = \sum_{k=1}^{K} \beta_{z,k}^{\star} B_{k}(t),\\
\end{split} \label{eq: function 2}
\ee
\vspace{-4ex}\\
To probabilistically cluster the $\bbeta_{x}$'s, we next let 
\vspace{-5ex}\\
\be
\begin{split}
	& z_{x} \sim \Mult(\bpi_{z}) = \Mult(\pi_{1},\dots,\pi_{z_{\max}}),\\
	&\bpi_{z} \sim \Dir(\alpha/z_{\max},\dots,\alpha/z_{\max}).
\end{split} \label{eq: function 2}
\ee 
\vspace{-4ex}\\
We next consider priors for the atoms $\bbeta_{z}^{\star}$. 
We let 
\vspace{-4ex}\\
\be
& \bbeta_{z}^{\star} \sim 
\MVN_{K}\{\bmu_{\beta,0},(\sigma_{a}^{-2}\bI_{K}+\sigma_{s}^{-2}\bP)^{-1}\}, \label{eq: random effects}
\ee
\vspace{-4ex}\\
where $\MVN_{K}(\bmu,\bSigma)$ denotes a $K$ dimensional multivariate normal distribution with mean $\bmu$ and covariance $\bSigma$ 
and $\bP = \bD\trans \bD$, where the $(K-1) \times K$ matrix $\bD$ is such that $\bD \bbeta$ computes the first order differences in $\bbeta$.
The model thus penalizes $\sum_{k=1}^K (\nabla \beta_{z,k}^{\star})^{2} = \bbeta \trans \bP \bbeta$, the sum of squares of first order differences in $\bbeta_{u}^{(i)}$ \citeplatex{eilers1996flexible}. 
The variance parameter $\sigma_{s}^{2}$ models the smoothness of the functional atoms, smaller $\sigma_{s}^{2}$ inducing smoother $f_{z}^{\star}(t)$'s. 
Additional departures from $\bmu_{\beta,0}$ are explained by the other variance component $\sigma_{a}^{2}$. 
We assign half Cauchy priors on the variance parameters as
\vspace{-4ex}\\
\bse
& \sigma_{s}^{2} \sim \HC (0, 1),~~~~~\sigma_{a}^{2} \sim \HC (0, 1).
\ese

\subsection{Functional Random Effects} \label{sec: random effects}
We allow different random effects $u_{C}^{(i)}(t)$ and $u_{I}^{(i)}(t)$ for correct (C) (when $d=s$) and incorrect (I) (when $d\neq s$) identifications, respectively, as \vspace{-4ex}\\
\bse
& u_{d,s}^{(i)}(t) = u_{C}^{(i)}(t)~~~~\text{when}~d=s,
~~~~~u_{d,s}^{(i)}(t) = u_{I}^{(i)}(t)~~~~\text{when}~d \neq s.
\ese 
\vspace{-4ex}\\
Suppressing the subscripts to simplify notation, 
we model the time-varying random effects components $u^{(i)}(t)$ as 
\vspace{-4ex}\\
\be
\begin{split}
	& \textstyle u^{(i)}(t) = \sum_{k=1}^{K} \beta_{u,k}^{(i)} B_{k}(t) = \bB(t) \bbeta_{u}^{(i)}, ~~~~~\\
	& \bbeta_{u}^{(i)} \sim 
	\MVN_{K}\{\bzero,(\sigma_{u,a}^{-2}\bI_{K}+\sigma_{u,s}^{-2}\bP)^{-1}\},~~~~~\\
\end{split} \label{eq: random effects}
\ee
\vspace{-2ex}\\
where $\bbeta_{u}^{(i)} = (\beta_{1,u}^{(i)},\dots,\beta_{K,u}^{(i)})\trans$ are subject-specific spline coefficients. 
We assign non-informative half-Cauchy priors on the variance parameters as
\vspace{-4ex}\\
\bse
& \sigma_{u,s}^{2} \sim \HC (0, 1),~~~~~\sigma_{u,a}^{2} \sim \HC (0, 1).
\ese

%\vskip 10pt
\section{Prior Hyper-parameters and Initialization} \label{sec: prior hyper-parameters}

The random effects of the inverse-probit mixed model are all initialized at zero. The variance and smoothing parameters are initially set to $0.1$ each. The location parameter of the prior on $\bbeta_{z}^{\star}$, $\bmu_{\beta,0}$ is set to $(1,\ldots,1)$. This choice of $\bbeta_{z}^{\star}$ would set the expected value of $\mu_{x}^{(i)}(t)$ to $1$, which is supported empirically. 
The value of the parameter $\alpha$ is set to $1$.

%\vskip 10pt
\section{Posterior Inference} \label{sec: post inference sm}

Posterior inference for the longitudinal drift-diffusion mixed  model, described in Section 3 in the main paper, 
is based on samples drawn from the posterior using an MCMC algorithm. 
The algorithm carefully exploits the conditional independence relationships encoded in the model 
as well as the latent variable construction of the model. 
Sampling the latent inverse-Gaussian distributed response times, in particular, greatly simplifies computation.

In what follows, $\bzeta$ denotes a generic variable that collects all other variables not explicitly mentioned, including the data points.
Also, $p_{0}$ will sometimes be used as a generic for a prior distribution without explicitly mentioning its hyper-parameters. 
The notation $x$ is used to abbreviate $(d^{\prime},s)$.   
The sampler for the inverse-probit mixed model of Section 3 iterates between the following steps.

\begin{enumerate}[leftmargin=0cm,itemindent=.5cm,labelwidth=\itemindent,labelsep=0cm,align=left]
	\item {\bf Sampling $\tau_{1:d_{0}}^{(i,l)}(t)$:} 
	Suppose the $i$-th individual selects the output tone $d$, in the $t$-th block, $l$-th trial, given the input tone $s$. 
	Then $\tau_{1}^{(i,l)}(t),\ldots, \tau_{d_{0}}^{(i,l)}(t)$ is generated as in Algorithm 1 (see Section 4) from the joint distribution of $\tau_{1}^{(i,l)}(t),\ldots, \tau_{d_{0}}^{(i,l)}(t)$ given $\mu_{1,s}^{(i)}(t), \ldots, \mu_{d_{0},s}^{(i)}(t)$, followed by an accept reject step. 
	
	\item {\bf Updating the components of fixed effects $f_{x}(t)$:} 
	\begin{enumerate}
		\item ~The latent variable $\bz_{x}$, 
		indicating the group identities of $\bbeta_{x}$, 
		follows multinomial distribution with $z_{\max}$ labels and probabilities
		$P(z_{x}=z|\bzeta),~z=1,\ldots, z_{\max}$ a posteriori.
		The probability $P(z_{x}=z|\bzeta)\propto \pi_{z}\times l_{z}$, where $l_{z}$ is the likelihood of $\bbeta_{x}$ evaluated at $\bbeta_{z}$.
		Let $L^{\star}$ be the set of all trials corresponding to input-output tones $x=(s,d)$ for $i$-th individual and $t$-th block, and $n_{x}^{(i)}(t)$ be the cardinality of $L^{\star}$.
		Furthermore, let $\tau_{x}^{(i)}(t)=\sum_{l\in L^{\star}} \tau_{x}^{(i,l)}(t)$, $\tau_{x}(t)=\sum_{i}\tau_{x}^{(i)}(t)$, and $n_{x}(t)=\sum_{i} n_{x}^{(i)}(t)$.
		A little algebra shows that the likelihood of $\bbeta_{x}$ is
		Gaussian with variance matrix $\bSigma_{\beta,x}=\left\{ \sum_{t} \tau_{x}(t) \bB(t)^T \bB(t) \right\}^{-1}$, and mean vector $\bmu_{\beta,x}=\bSigma_{\beta,x}  \left\{ \sum_{t} \bB(t)M_{x}(t) \right\}$, where $M_{x}(t)=2 n_{x}(t)- \sum_{i} u_{x}^{(i)}(t) \tau_{x}^{(i)}(t) $.
		Therefore, $l_{z}$ is the Gaussian likelihood with mean $\bmu_{\beta,x}$ and variance $\bSigma_{\beta,x}$, evaluated at $\bbeta_{z}$.
		
		\item ~Let $N_{z}=\sum_{x} 1(z_{x}=z)$, $z=1,\ldots,z_{\max}$, where $1(\cdot)$ is the indicator function. Then the conditional posterior of $\bpi_{z}$ is Dirichlet with parameters $\alpha/z_{\max}+N_{1}, \ldots, \alpha/z_{\max}+N_{z_{\max}}$.
		
		\item ~The full conditional posterior distribution of the coefficient atoms $\bbeta_{z}^{\star}$ is Gaussian with variance-covariance matrix $\bSigma_{\beta,z}^{\star}$ and $\bmu_{\beta,z}^{\star}$, where 
		$\bSigma_{\beta,z}^{\star,-1}=  \sum_{x:z_{x}=z}   \bSigma_{\beta,x}^{-1}+\bSigma_{\beta,0}^{-1}$, and  
		$\bmu_{\beta,z}^{\star}=\bSigma_{\beta,z}^{\star}\left[ \sum_{x:z_{x} =z} \bSigma_{\beta,x}^{-1} \bmu_{\beta,x}+\bSigma_{\beta,0}^{-1} \bmu_{\beta,0} \right],$
		where $\bSigma_{\beta,0}^{-1}=\left( \sigma_{a}^{-2} \bI_{K}+\sigma_{s}^{-2}\bP \right)$.
	\end{enumerate}
	
	\item {\bf Updating the components of random effects:} We use the generic notation $U$ to indicate the correct ($C$, i.e., $d=s$) or incorrect ($I$, i.e, $d\neq s$) cases.
	Define $\tau_{U}^{(i)}(t)=\sum_{x:x\in U} \tau_{x}^{(i)}(t)$, $n_{U}^{(i)}(t)=\sum_{x:x\in U} n_{x}^{(i)}(t)$, $ f\tau_{U}^{(i)}(t) =\sum_{x:x\in U} \tau_{x}^{(i)}(t) f_{x}(t)$, $\bSigma_{U,0}^{-1}=\sigma_{U,a}^{-2} \bI_{K} + \sigma_{U,s}^{-2} \bP $, and $\bSigma_{U}^{(t)-1}= \sum_{t} \tau_{U}^{(i)}(t) \bB(t)^{T} \bB(t)$.
	The conditional posterior of $\bbeta_{U}^{(i)}$ is Gaussian with covariance $\bSigma_{U,\mathrm{post}}^{(i)}=\left( \bSigma_{U,0}^{-1}+\bSigma_{U}^{(i)-1}\right)^{-1}$, and location parameter $\bmu_{C}^{(i)}=\bSigma_{U,\mathrm{post}}^{(i)}\bSigma_{U}^{(i)-1}\left[\sum_{t}\left\{2 n_{U}^{(i)}(t)-f\tau_{U}^{(i)}(t) \right\} \bB(t)  \right]$, respectively.
	
	\item {\bf Updating the precision and smoothing parameters:}
	The precision and smoothness parameters involved in the fixed effects part are $\sigma_{a}^{2}$ and $\sigma_{s}^{2}$, and those involved in the random effects part are $\sigma_{U,a}$ and $\sigma_{U,s}^{2}$, $U=C,I$. We update these variance components using Metropolis-Hastings algorithm with log-normal proposal distributions centered on the previous sample values.

	\item {\bf Estimation of probability:}
	For each $(s,i,t)$, we calculate the probability of selecting the $d$-th response in the following way:
	Let $g\{\cdot \mid \mu_{d^{\prime},s}^{(i)}(t) \}$ be the pdf of inverse Gaussian distribution of the form (1) with parameters $\delta_{s}=0$, $b_{d^{\prime},s}=2$ and $\mu_{d^{\prime},s}=\mu_{d^{\prime},s}^{(i)}(t)$.
	We generate $M=2000$ independent samples $\btau_{m}=\left[\tau_{1,m},\ldots,\tau_{d_{0},m} \right]^{T}$, $m=1,\ldots,M$, where $\tau_{d^{\prime},m}$ is generated independently from $g\{\cdot \mid \mu_{d^{\prime},s}^{(i)}(t)\}$. 
	Among these $M$ independent samples, the proportion of occurrences of $\{\tau_{d,m}\leq \wedge_{d^{\prime}=1:d_{0}} \tau_{d^{\prime},m}\}$ is considered as the estimated probability of selecting $d\th$ response.

\end{enumerate}

The results reported in this article are all based on $5,000$ MCMC iterations with the initial $2,000$ iterations discarded as burn-in.
The remaining samples were further thinned by an interval of $5$.
We programmed in \texttt{R}. 
The codes are available as part of the supplementary material. 
A `readme' file, providing additional details for a practitioner, is also included in the supplementary material.
In all experiments, the posterior samples produced very stable estimates of the population and individual level parameters of interest. 
MCMC diagnostic checks were not indicative of any convergence or mixing issues.

\section{MCMC Diagnostics}\label{sec: MCMCdiag}
This section presents some convergence diagnostics for the MCMC sampler described in the main manuscript. 
The results presented here are for the PTC1 data set. 
Diagnostics for the simulation experiments and the benchmark data were similar and hence omitted.

\begin{figure}[ht!]
	\centering
	\includegraphics[scale=.65,trim=0cm 1cm 0cm 1cm, clip=TRUE]{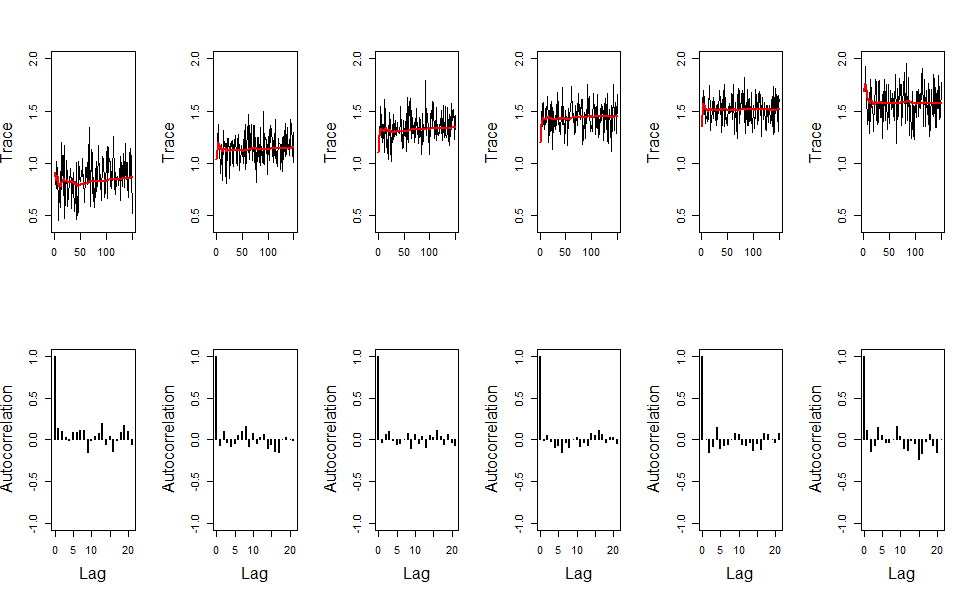}
	\caption{\baselineskip=10pt\small Analysis of PTC1 data:  Trace plots (top) and auto-correlation plots (bottom) of the individual drift rates $\mu_{1,1}^{(1)}(t)$ corresponding to the success categorization of tone $T_{1}$ evaluated at each of the training blocks. In each panel, the solid red line shows the running mean.
		Results for other drift parameters were very similar.}
	\label{fig: PTC1_MCMC}
\end{figure}

Figure \ref{fig: PTC1_MCMC} shows the trace plots and auto-correlation of some individual level parameters at different training blocks. These results are based on the MCMC thinned samples. As these figures show, the running means are very stable and there seems to be no convergence issues. Additionally, the Geweke test \citep{Geweke1992Evaluating} for stationarity of the chains, which formally compares the means of the first and last part of a Markov chain, was also performed. If the samples are drawn from the stationary distribution of the chain, the two means are equal and Geweke statistic has an asymptotically standard normal distribution. The results of the test, reported in Table \ref{PTC1_Geweke_stat}, indicate that convergence was satisfactory.

\begin{table}[ht]
	\centering {\small
		\begin{tabular}{|c|c|c|c|c|c|c|}\hline
			& $t=1$ & $t=2$ & $t=2$ & $t=2$ & $t=2$ & $t=2$ \\ \hline 
			Geweke statistics  & -1.233 &  -0.392 &  -0.678 &  -0.136 & 0.440 & 0.339 \\
			$p$-value & 0.217 & 0.695 & 0.498 & 0.892 & 0.660 & 0.734 \\ \hline
		\end{tabular}
		\caption{Geweke statistics and associated p-values assessing convergence of the of the individual level drift parameters $\mu_{1,1}^{(1)}(t)$ corresponding to the success categorization of tone $T_{1}$ evaluated at each of the training blocks. Results for other drift parameters were very similar.
		}
		\label{PTC1_Geweke_stat}}
\end{table}

\newpage
\section{Analysis of Benchmark Data}\label{sec: ABMD}

\paragraph{Description of the data.}
The data set we consider next is a multi-day longitudinal speech category training study reported previously in \citelatex{reetzke2018tracing} and analyzed previously in \citelatex{paulon2021bayes}. 
In this study,   
$n = 20$ participants were trained to learn $4$ tones, namely, 
high-level (T1), low-rising (T2), low-dipping (T3), or high-falling (T4) tone, respectively. 
The trials were administered in blocks, each comprising $40$ categorization trials.
Participants were trained across several days, with five blocks on each day. 
On each trial, participants indicated the tone category they heard via button press on a computer keyboard. 
Following the button press, they were given corrective feedback. 
The data consist of tone responses and associated response times for different input tones for the $20$ participants. 
We focus here on the first two days of training (10 blocks in total) 
as they exhibited the steepest improvement in learning as well as the most striking individual differences relative to any other collection of blocks.

\begin{figure}[ht!]
	\centering
	\includegraphics[scale=.45]{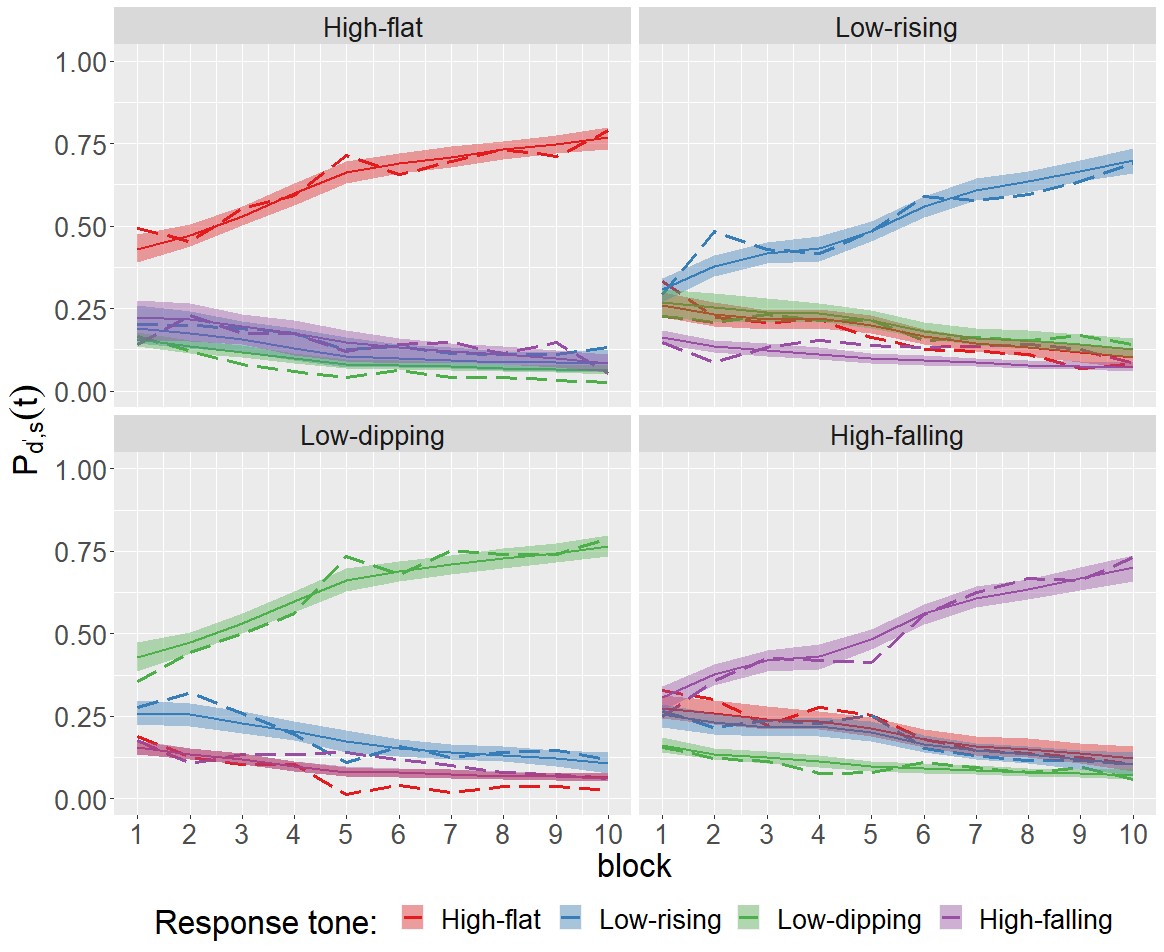}
	\caption{\baselineskip=10pt\small
		Results for the benchmark data: Estimated probability trajectories compared with average proportions of times an input tone was classified into different tone categories across subjects (in dashed line).
		High-flat tone responses are shown in red; low-rising in blue; low-dipping in green; and high-falling in purple. 
	}
	\label{fig: RD1_estimated_prob}
\end{figure}

\paragraph{Analysis.} We first demonstrate the performance of the proposed method in estimating the probabilities associated with different $(d,s)$ pairs. 
Figure \ref{fig: RD1_estimated_prob} shows the $95\%$ credible intervals for the estimated probabilities for different input tones  
along with the average  proportions of times an input tone was classified into different tone categories across subjects.

Observe that, except in situations with a very small number of data points
the $95\%$  credible intervals include the empirical probabilities. 
Further, the estimated credible region is narrow enough implying high precision of the inference.

\begin{figure}[ht!]
	\centering
	\includegraphics[scale=.65,trim=0cm 2.5cm 2cm 2.5cm,clip=TRUE]{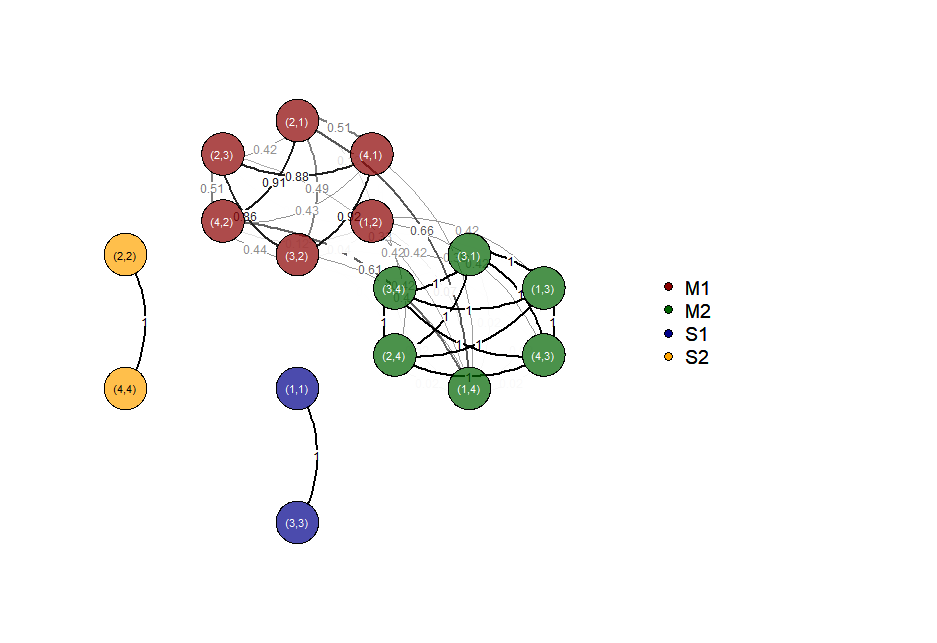}
	\caption{\baselineskip=10pt\small
		Results for the benchmark data: Network plot of similarity groups showing the intra and inter-cluster similarities. 
		Each node is associated with a pair indicating the input-response tone category $(s,d)$. 
		The number associated with each edge indicates the proportion of times the pair in the two connecting nodes appeared in the same cluster after burning.
	}
	\label{fig: Benchmark_data_network}
\end{figure}

Next, consider the clustering results.
We obtained two clusters each in pairs of success combinations $(d=s)$ and in the wrong allocations $(d\neq s)$. 
The clusters of success combinations are 
{\small $S_{1}=\{(1,1),(3,3)\}$} and {\small $S_{2}=\{(2,2),(4,4)\}$}, 
and that in wrong allocations are {\small$M_{1}=\{(1,2),(2,1),(2,3),(3,2),(4,1),(4,2)\}$}, and {\small $M_{2}=\{(1,3),(1,4),(2,4),(3,1),(3,4),(4,3) \}$}. 
The network plot in Figure \ref{fig: Benchmark_data_network} shows the stability of the clusters over the MCMC iterations.

From an overall perspective, 
the trajectory of `High-level' ($T_{1}$) and `Low-dipping' ($T_{3}$) are similar with two wrong allocations from $M_{2}$ and one from $M_{1}$, 
and that of `Low-rising' ($T_{2}$) and `High-failing' ($T_{4}$) are similar with two wrong allocations from $M_{1}$ and one from $M_{2}$. 
These similarities in the overall trajectories of {\small $\{T_{1},T_{3}\}$} and {\small $\{T_{2},T_{4}\}$} were also noted by \citelatex{paulon2021bayes}.  

Next, we consider the estimation of the underlying drift parameters $\mu_{d^{\prime},s}^{(i)}(t)$. 
Due to the identifiability constraints, 
the estimates of $\mu_{d^{\prime},s}^{(i)}(t)$ can only be observed on a relative scale. 
Figure \ref{fig: population_effect} shows the posterior mean trajectories and associated $95\%$ credible intervals for the projected drift rates estimated by our method for different combinations of $(d^{\prime},s)$. 
In comparison with the previous analysis of \citelatex{paulon2021bayes}, 
the trajectories of our estimated drift rates show significant similarity throughout.

\afterpage{
	\begin{figure}[ht!]
		\centering
		\includegraphics[scale=.45, trim=2cm 0.25cm 1cm 0.25cm]{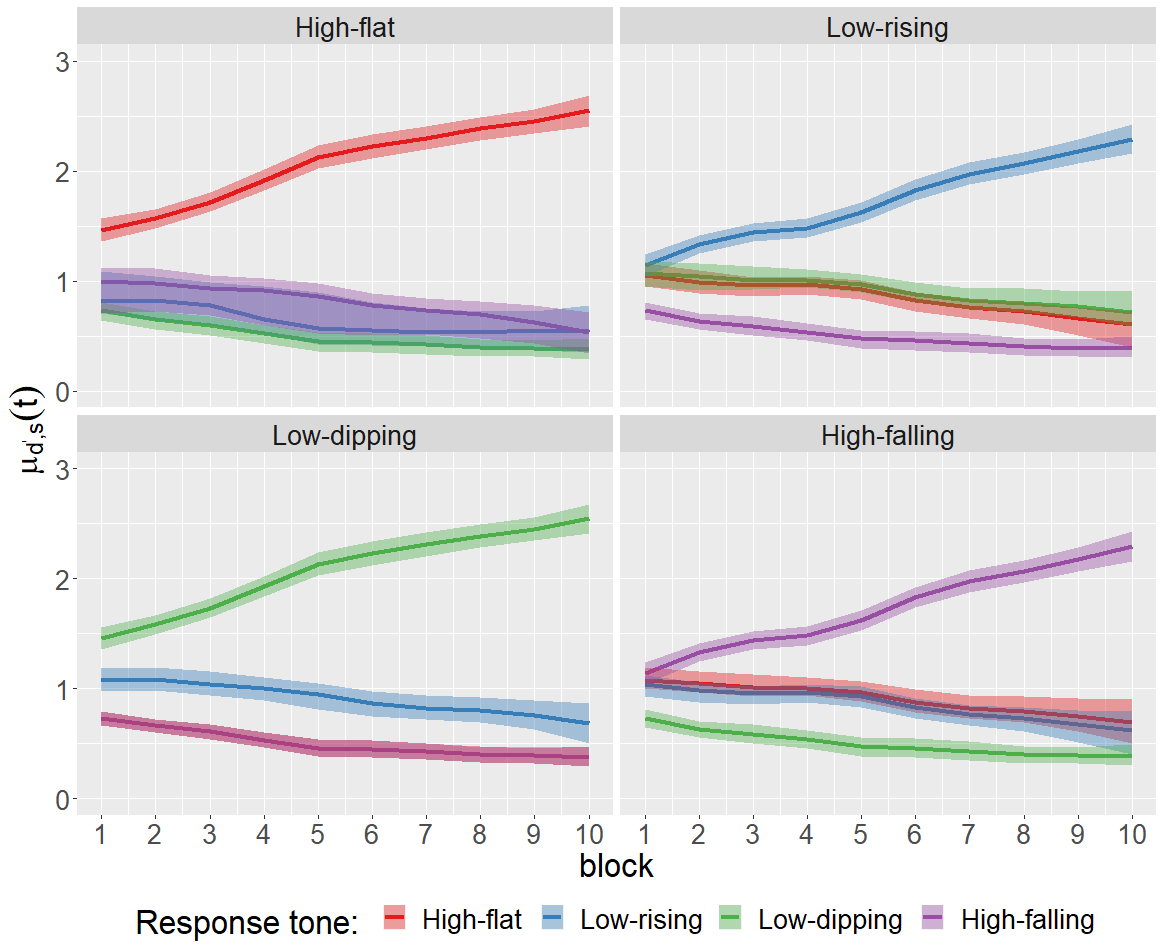}
		\caption{\baselineskip=10pt\small Results for the benchmark data: Estimated posterior mean trajectories of the population level drifts $\mu_{d^{\prime},s}(t)$ 
			for the proposed model. 
			The shaded areas represent the corresponding $95\%$ pointwise credible intervals.
			Parameters for the high-flat tone response category are shown in red; low-rising in blue; low-dipping in green; and high-falling in purple.}
		\label{fig: population_effect}
	\end{figure}
}

Figure \ref{fig: random_effect} shows the posterior mean trajectories and associated $95\%$ credible intervals 
for the drift rates $\mu_{d^{\prime},s}^{(i)}(t)$ for the different correct combinations $(d^{\prime},s)$ with $d^{\prime}=s$ for two participants 
- the one with the best accuracy averaged, and the one with the worst accuracy averaged across all blocks.
For the well-performing participant, the drift trajectories increase rapidly and
for the poorly performing candidate, on the other hand, the drift trajectories increase very slowly.
Once again, 
in spite of the limitation of inferring on a relative scale,
the relative differences of the best and worst performing participants across blocks show great similarity with the inference of \citelatex{paulon2021bayes}.

\afterpage{
	\begin{figure}[ht!]
		\centering
		\includegraphics[scale=.45]{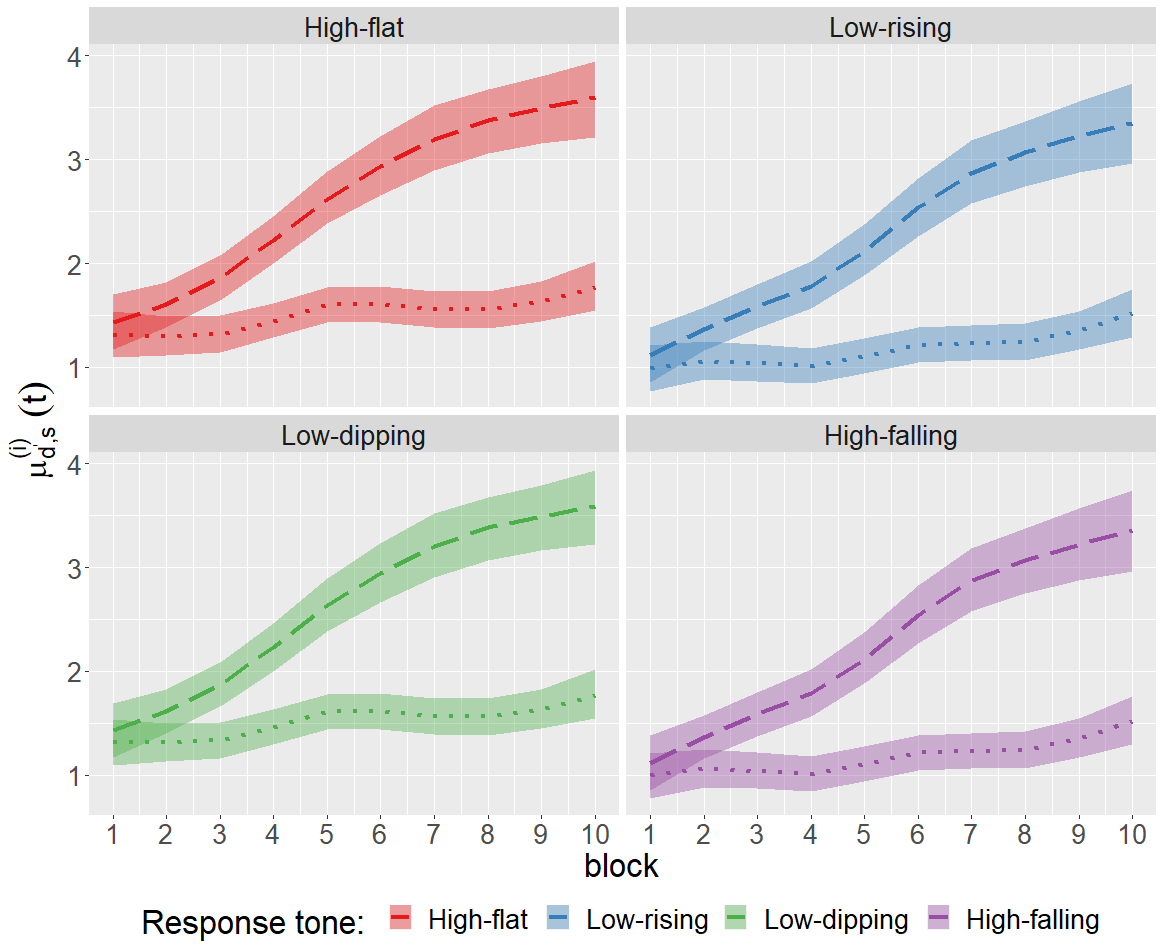}
		\caption{\baselineskip=10pt\small 
			Results for the benchmark data: 
			Estimated posterior mean trajectories for individual specific drifts $\mu_{d^{\prime},s}^{(i)}(t) = \exp \{ f_{d^{\prime},s}(t) + u_{C}^{(i)}(t) \}$ 
			for correct identification $(d^{\prime}=s)$ for two different participants 
			- one performing well (dashed line) and one performing poorly (dotted line). 
			The shaded areas represent the corresponding $95\%$ point-wise credible intervals.
			Parameters for the high-flat tone response category are shown in red; low-rising in blue; low-dipping in green; and high-falling in purple.}
		\label{fig: random_effect}
\end{figure}}

\newpage
\section{Rand and Adjusted Rand Indices}\label{sec: Rand}
\paragraph{Rand Index.} Given a set of $n$ objects $S=\{s_{1},\ldots,s_{n}\}$, let ${\bf U}=\left\{U_{1}, \ldots, U_{R}\right\}$ and ${\bf V}=\left\{V_{1}, \ldots, V_{C}\right\}$ represent two different partitions of the objects in $S$ such that $\cup_{i=1}^{R} U_{i}=S=\cup_{j=1}^{C} V_{j}$ and $U_{i} \cap U_{i^{\prime}}=\emptyset=V_{j} \cap V_{j^{\prime}}$ for $1 \leq i \neq i^{\prime} \leq R$ and $1 \leq j \neq j^{\prime} \leq C$. Rand index estimates the similarity between the allocations of $S$ in ${\bf U}$ and ${\bf V}$.

Let $a$ be the number of pairs of objects that are placed in the same partition in ${\bf U}$ and the same partition in ${\bf V}$, and $b$ be the number of pairs of objects that are in different partitions of ${\bf U}$, as well as in different partitions of ${\bf V}$.
Here $a$ and $b$ can be interpreted as agreements in ${\bf U}$ and ${\bf V}$, and the total number of pairs is $\binom{n}{2}$.
The Rand index \citeplatex{Rand1971Objective} is 
\vskip-2ex
$${\tt RI}= (a+b)/\binom{n}{2}.$$ 
\vskip-1ex

\noindent The Rand index lies between 0 and 1. When the two partitions agree perfectly, the ${\tt RI}$ takes the value 1.

\paragraph{Adjusted Rand Index.} 
The expected value of the Rand index of two random partitions does not take a constant value. The adjusted Rand index \citeplatex{Hubert1985Comparing} assumes
generalized hypergeometric distribution as the model of randomness, and makes a base and scale change of the quantity $(a+b)$, defined above, so that the resultant quantity is bounded by $[-1,1]$ and has expected value $0$ under completely random allocation.  

Let $n_{i,j}$ be the number of object that are both in $i\th$ partition of ${\bf U}$ and $j\th$ partition of ${\bf V}$, $n_{i}$ and $n_{j}$ be the total number of components in $i\th$ partition of ${\bf U}$, and $j\th$ partition of ${\bf V}$, respectively.  

The expression $a+d$ can be simplified to a linear transformation of $\sum_{i, j}\binom{n_{i,j}}{2}$. Further, under the generalized hypergeometric model, it can be shown that

\vskip-1ex
{\small$$ \eE\left[\sum_{i,j}\binom{n_{i j}}{2}\right]=\left[\sum_{i}\binom{n_{i}}{2} \sum_{j}\binom{n_{j}}{2}\right] /\binom{n}{2}.$$}
\vskip-1ex

\noindent Therefore, scaled the difference of linear transformed $(a+b)$ and its expectation is the adjusted Rand index, defined as:
$$ {\tt ARI}=
\frac{\sum_{i, j}\binom{n_{i,j}}{2}-\left[\sum_{i}\binom{n_{i}}{2} \sum_{j}\binom{n_{i,j}}{2}\right] /\binom{n}{2}}{\frac{1}{2}\left[\sum_{i}\binom{n_{i}}{2}+\sum_{j}\binom{n_{j}}{2} \right]-\left[\sum_{i}\binom{n_{i}}{2} \sum_{j}\binom{n_{j}}{2}\right] /\binom{n}{2}}.
$$
The expected value of ${\tt ARI}$ index is zero and the range is $[-1,1]$. 
Like the {\tt RI}, the {\tt ARI} also takes the value $1$, when the two partitions agree perfectly.

\bibliographystylelatex{natbib}
\bibliographylatex{Categorical,Diffusion,FDA_LDA,HMM,HOHMM,MCMC_Latent_Var_Models,Projection,neuro}

\end{document}